\newcommand{\vdate}{July 1993}
\newcommand{\epem}{$\mbox{e}^+ \mbox{e}^-$}
\newcommand{\reff}[1]{(\ref{#1})}
\newcommand{\nonu}{\nonumber\\}
\newcommand{\porder}[1]{\mbox{${\cal O}(#1)$}}
\newcommand{\yqi}{\mbox{$y_{qi}$}}
\newcommand{\yqg}{\mbox{$y_{qg}$}}
\newcommand{\yig}{\mbox{$y_{ig}$}}
\newcommand{\yqqb}{\mbox{$y_{q{\overline q}}$}}
\newcommand{\yqbi}{\mbox{$y_{{\overline q} i}$}}
\newcommand{\qbar}{\mbox{${\overline q}$}}
\newcommand{\iqbar}{\mbox{${\scriptstyle\overline q}$}}
\newcommand{\dilog}{{{\cal L}_2}}
\newcommand{\spence}{{\cal S}}
\newcommand{\lln}{l}
\newcommand{\cnumber}[0]{\mbox{\boldmath$C$}}

\newcommand{\munit}[0]{\mbox{\bf 1}}
\newcommand{\tr}[0]{\mbox{tr}}
\newcommand{\GeV}{\mbox{GeV}}
\newcommand{\SH}{{S_{\!\scriptscriptstyle H}}}
\newcommand{\XB}{{x_{\!\scriptscriptstyle B}}}
\newcommand{\SS}{\scriptscriptstyle}
\newcommand{\grkl}{{\raisebox{-0.2cm}{$>$} \atop \raisebox{0.2cm}{$<$}}}
\documentstyle[11pt]{article}

\newlength{\dinwidth}
\newlength{\dinmargin}
\begin{document}
\begin{titlepage}

\renewcommand{\thefootnote}{\fnsymbol{footnote}}
\vdate\hfill
\hspace*{\fill}
\begin{minipage}[t]{2.7cm}
LBL--34147
\end{minipage}
\begin{center}

\vspace{8mm}
{\LARGE
Next-to-Leading Order QCD Corrections\\
to
Jet Cross Sections and
Jet Rates\\
in Deeply Inelastic
Electron Proton
Scattering\footnote[1]{{\em
This work was supported by the Director, Office of Energy
Research, Office of High Energy and Nuclear Physics, Division of High
Energy Physics of the U.S. Department of Energy under Contract
DE-AC03-76SF00098 and the Bundesministerium f\"ur Forschung und
Technologie under contract number 05~5AC~91~P.}}\footnote[2]{
{\em Revised version.}}\\
}
\vspace{1cm}
{\large
Dirk Graudenz$\;$\footnote[3]{{\em supported by Max Kade Foundation,
New York}}\footnote[4]{{\em Address after January 1, 1994:
CERN, Theoretical Physics Division, CH-1211 Geneva,
Switzerland}}\footnote[5]{{\em E-mail addresses:
graudenz @ theorm.lbl.gov, GRAUDENZ @ LBL.bitnet, I02GAU @ DHHDESY3.bitnet}} \\
Theoretical Physics Group\\
LBL / University of California\\
1 Cyclotron Road\\
Berkeley, CA 94720, USA\\
}
\end{center}

\vspace{1cm}

\begin{abstract}
Jet cross sections in deeply inelastic
scattering in the case of transverse
photon exchange
for the production of
(1+1) and (2+1) jets are calculated in next-to-leading order QCD
(here the `+1' stands for the target remnant jet, which is included in the jet
definition for reasons that will become clear in the main text).
The jet definition scheme is based on a modified JADE cluster algorithm.
The calculation of the (2+1) jet cross section
is described in detail.
Results for the virtual corrections as well as for the
real initial- and final state corrections are given explicitly.
Numerical results are stated for jet cross sections as well as
for the ratio
$\sigma_{\mbox{\small (2+1) jet}}/\sigma_{\mbox{\small tot}}$ that can
be expected at E665 and HERA.
Furthermore the scale ambiguity of the calculated
jet cross sections is studied
and different parton density parametrizations are compared.
\end{abstract}

\end{titlepage}

\renewcommand{\thefootnote}{\arabic{footnote}}
\setcounter{footnote}{0}

\renewcommand{\thepage}{\roman{page}}
\setcounter{page}{2}
\mbox{ }

\vskip 1in

\begin{center}
{\bf Disclaimer}
\end{center}

\vskip .2in

\begin{scriptsize}
\begin{quotation}
This document was prepared as an account of work sponsored by the United
States Government.  Neither the United States Government nor any agency
thereof, nor The Regents of the University of California, nor any of their
employees, makes any warranty, express or implied, or assumes any legal
liability or responsibility for the accuracy, completeness, or usefulness
of any information, apparatus, product, or process disclosed, or represents
that its use would not infringe privately owned rights.  Reference herein
to any specific commercial products process, or service by its trade name,
trademark, manufacturer, or otherwise, does not necessarily constitute or
imply its endorsement, recommendation, or favoring by the United States
Government or any agency thereof, or The Regents of the University of
California.  The views and opinions of authors expressed herein do not
necessarily state or reflect those of the United States Government or any
agency thereof of The Regents of the University of California and shall
not be used for advertising or product endorsement purposes.
\end{quotation}
\end{scriptsize}

\vskip 2in

\begin{center}
\begin{small}
{\it Lawrence Berkeley Laboratory is an equal opportunity employer.}
\end{small}
\end{center}

\newpage
\renewcommand{\thepage}{\arabic{page}}

\setcounter{page}{1}

\newpage

\section{Introduction}
Recent results from E665 and
HERA show that events with a clear jet-like structure
are present in deeply inelastic electron proton scattering
\cite{1,2}.
With sufficient luminosity it should therefore be possible to study
jet cross sections
to use them for a test of QCD and
an independent determination of its fundamental parameter
$\Lambda_{\mbox{\small QCD}}$.
Jet cross sections may even be useful to
extract some information on the gluon density at small $x$, because
the gluon density is important in
(2+1)\footnote{The target remnant is counted as a jet, so ``2+1''
stands for the production of 2~partons in the hard QCD process,
possibly accompanied by additional soft or collinear partons.}
jet production.

Because of the strong scale dependence of fixed order cross sections
calculated in perturbative QCD,
the possibility of a large size of the corrections
and because a determination of $\alpha_s$ must be based on
a next-to-leading (NLO) order cross section,
the calculation of higher order
corrections
is well motivated.
Since now experimental results are available, it is worthwhile
to give a detailed account of the
technical problems of the calculation
of the (2+1) jet cross section.
A second goal of the present work is to
study the jet cross sections in detail numerically.
In deeply inelastic scattering, the \porder{\alpha_s} corrections
to the \porder{\alpha_s^0} Born term are well known
(see \cite{3,4,5,6,7,8,9}).
In addition, the \porder{\alpha_s^2} Born terms for the production
of (3+1) jets have been calculated \cite{10,11}.
In this paper the calculation of the cross section for the
production of (2+1) jets to \porder{\alpha_s^2} in the case of transverse
photon exchange\footnote{Here,
the cross section for transverse photon exchange
is defined by the helicity cross section obtained by a contraction of the
hadron tensor with the metric tensor $(-g_{\mu\nu})$.} is described.
For more details of the calculation see \cite{12,13}.
The contributions with a transverse exchanged photon dominate the
cross section, as is shown in Section \ref{nr}.
More recently, the other parity-conserving helicity
cross sections have been calculated based on \cite{12}, see
\cite{14}.

One of the main features of hadronic events in high-energy collisions
is the pronounced jet-like structure. These jets are attributed to the
production of partons in the fundamental QCD process \cite{15}.
Due to the presence of infrared divergences in QCD, a suitable prescription
has to be given in order to define the finite parts that
arise from divergent terms after the singularities between
real and virtual corrections have been cancelled. Such a prescription
is related to the jet definition that is used on the parton level
to calculate jet cross sections.
In Section \ref{old2}
jet definitions in deeply inelastic scattering are discussed and the
jet definition scheme based on a modification of the
JADE cluster algorithm that
is used in this calculation is defined.

In Section \ref{new3} the calculation of the (1+1) jet cross section
in next-to-leading order is reviewed
and the results for the (2+1) jet Born terms are stated.
In Section \ref{old3}
the results for the virtual
corrections to (2+1) jet production are given.
In Sections \ref{old4} and
\ref{old5} the calculation of the real corrections to (2+1) jet production
is described (separated according to singularities in the final and
initial state). The sum of virtual and real corrections gives the finite
jet cross section after renormalisation of the parton densities. The
flavour factors are listed in Section
\ref{old6}. In Section \ref{old7}
the numerical results for jet cross sections and for
the ratio
$\sigma_{\mbox{\small (2+1) jet}}/\sigma_{\mbox{\small tot}}$ are presented.
The dependence of the jet cross sections on the
renormalisation and factorization scales, the dependence on the
jet definition scheme and the dependence on the parametrization of the
parton densities is also studied.
The appendix contains
explicit results for massless 1-loop tensor structure integrals,
phase space integrals
and the virtual and real corrections.

\section{Jet Cross Sections}
\label{old2}
Jets must be defined in terms of
experimentally observable quantities. An experimental event is characterised
by the energies and momenta of the outgoing particles. Since at high
energies the multiplicity of the events is large and since presently
there is no practical
way (based on QCD) to describe hadron dynamics on the level of
observable particles, one has to try to extract information from experimental
data in a form that can be compared with theoretical results
from perturbative QCD.
At \epem-colliders it has been observed that outgoing hadrons very often
appear as clusters of particles concentrated
in a small cone in momentum space. These
clusters were called jets. Given an experimental event and a
resolution parameter $c$ (the ``jet cut''), a suitable algorithm is applied
to the event giving the number of jets of the event and the particles
associated with each of the jets. Therefore, the algorithm that is used
{\em defines} what is meant by a ``jet''. To compare experiment and
theory, one must use the same algorithm in theoretical calculations.
One should expect that experiment and theory are comparable as long as
the same algorithm is used in the experimental analysis and in the
theoretical calculation.
Of course, the problem is that in realistic events the final state consists of
hadrons, whereas in theoretical calculations (based on QCD) the outgoing
``particles'' are partons. Therefore the
crucial hypothesis is that jets on the hadron
level and jets on the parton level can be identified.

The first jet definition that has been given is that of Sterman and
Weinberg \cite{15}. It is based on cones in momentum space defined by an
opening angle $\delta$ and an energy fraction $\epsilon$. This definition,
which is well suited for \epem-annihilation for a  small number of jets
becomes complicated if a larger number of jets is produced. In addition,
it is not Lorentz invariant (this is not a problem in the case of
\epem-annihilation, since here the CM system is a unique frame of
reference). Later, another type of algorithm was proposed, the
cluster algorithm first used by the JADE collaboration \cite{16}.
This algorithm combines succesively two particles into a jet, if their
invariant mass squared $s_{ij}$ is smaller than a fraction $c$ of a typical
mass scale :
\begin{equation}
s_{ij}\le c M^2.
\end{equation}
In \epem-annihilation, $c$ is of the order of $10^{-2}$, and $M^2$ is
set to $Q^2$, the total invariant mass of the hadronic final state.

For hadron colliders, jet definitions in terms of ``cones'' in rapidity and
azimuthal angle are favoured (UA1-type algorithms).
Such a jet definition singles out a
particular axis (namely, the direction of the two colliding beams).
In the case of
pp-events
the incoming partons are assumed to have only small
transverse momentum, and therefore the situation is symmetric with respect to
this particular axis.

The situation is quite different for electron-proton colliders. Here the
interaction is mediated by the exchange of a virtual photon with momentum
$q$ (and $Q^2:=-q^2>0$). This photon hits a proton with momentum $P$.
Therefore, the interaction should be described in the CM system with
$\vec{P}+\vec{q}=\vec{0}$. This system varies from event to event, and this is
the reason why a Lorentz invariant jet definition should be used
\cite{11}.

Before a suggestion for a suitable jet algorithm in
eP-scattering is given,
one should have a closer look at the target remnant.
An interesting question is: ``Is it reasonable to include the target jet
in a jet analysis?''. One should consider the process in fig.~\ref{irad1}.
%
%
This Feynman
graph describes initial state radiation of a gluon with momentum $p_2$ from
the initial quark line with momentum $p_0$. It is assumed that the gluon is
emitted collinearly with a large energy
in the direction of the incoming quark causing a  strong
enhancement of the cross section for this process because of the pole
of the quark propagator. Since all partons are assumed to be massless
the gluon will go in the same direction as the target remnant. Under
the assumption that ``parton jets'' roughly correspond to ``hadron jets'',
there will be no possibility to disentangle the hadrons
from the debris of the proton and those coming from the fragmentation
of the gluon. To be consistent, one therefore should define ``parton jets''
in the following way: If an outgoing gluon can be separated from the
remnant by a suitable condition (e.g. invariant mass), the gluon and the
remnant are considered to be two jets with momenta $p_2$ and $p_r$.
If the gluon and the remnant cannot
be separated, they count as one jet whose momentum  $p_* = p_2+p_r$
is the sum of the
gluon momentum and the momentum of the remnant.
In an experiment, however, one cannot measure the momentum of the remnant
directly
since most of the hadrons from this jet are lost in the beam pipe.
The momentum of the target remnant jet must therefore be determined
in an indirect way.
The jet algorithm should also have the property that all collinear
singularities
are treated in such a way that they factorize the corresponding Born term. This
allows for a process independent definition of the renormalised scale dependent
parton densities.

For the experimental analysis it is proposed to use a modified JADE
cluster algorithm (mJADE algorithm) consisting of two steps (see also
\cite{11}):
\newline
(1) Define a {\em precluster} of longitudinal momentum (in the direction of the
beam pipe) $p_r$ that is given by the missing longitudinal momentum
of the event.
\newline
(2) Apply the JADE cluster algorithm to the set of momenta
\begin{equation}
\{p_1, p_2, \ldots, p_n, p_r\},
\end{equation}
 where $p_1, p_2, \ldots, p_n$ are the
momenta of the visible hadrons in the detector and $p_r$ is the momentum
of the precluster.

It remains to define the order of magnitude of the jet cut and
the mass scale $M^2$ to be used in the mJADE algorithm.
The jet cut should be such that it is small enough to ensure
properly separated jets in the detector (otherwise the jets
coming from the real corrections could become too broad), but large
enough to avoid large logarithms that could spoil a fixed order result in
perturbation  theory. One can think of $cM^2$ as a new mass scale that is
needed to specify the cross section completely. In a deeply inelastic
process there are several mass scales given by an event, namely the virtuality
$Q^2$ of the exchanged vector boson, the invariant mass $W^2$ of the
hadronic final state and, in general, several invariant masses
$p_{i\perp}^2$ from transverse momenta of outgoing particles.
Therefore large logarithms are expected if the quotient of any two of these
scales becomes small (or large). So one should avoid kinematical regions
where this could happen. Because of the jet cut
an additional scale $cM^2$ enters the calculation,
where one could, for example, use
$Q^2$, $W^2$ or some $p_{i\perp}^2$ as $M^2$. Since it is suggested to include
the target remnant in the jet analysis, it is natural to use the scale
$M^2=W^2$ in the jet algorithm (analogous to the situation in
\epem-annihilation). If the scales $p_{i\perp}^2$ are omitted from the
discussion,
the scales $Q^2=\SH y \XB$, $W^2=\SH y (1-\XB)$
and $cW^2=c \SH y (1-\XB)$ are relevant. Here $\sqrt{\SH}$ is the total CM
energy of the collider,
$\XB=Q^2/2Pq$ is the Bjorken variable and $y$ is the usual lepton
variable $y=Pq/Pk$, with P the proton momentum and k the momentum of
the incoming electron.

This is not the only reasonable choice
for the mass scale used in the
the jet definition. In fact, in
Section \ref{new3}
it is pointed out that a scale like $Q^2$ or
$\left(W^\alpha Q^\beta \sqrt{\SH y}^{1-\alpha-\beta})\right)^2$
with parameters
$\alpha,\;\beta$ may be reasonable if
the parton densities have to be probed at small $x$. In contrast,
the considerations here
are based on the simple observation that the observable invariants (including
those with the remnant jet) sum up to $W^2$, and it is somehow natural to use
this particular scale. For another possible jet definition in
eP scattering see \cite{17}.

In the cross section
large logarithms of $Q^2/W^2\approx \XB$, $c W^2/W^2=c$ and
$cW^2/Q^2\approx c/\XB$ (the approximation of small $\XB$ is made since
most events are expected in this region) are expected. For cuts of the
order of 0.01 the logarithms in c should be comparable to those
encountered in
\epem-annihilation because of a similar structure of the matrix elements.

Berger and Nisius have studied the effect of the inclusion of the target
remnant in the jet analysis \cite{18} by using the Monte Carlo generator
LEPTO5.2. They come to the conclusion that the correlation of the number
of jets on the parton level and the number of jets after fragmentation
is much stronger if the remnant jet is included. For more details, see
\cite{11}.

\section{(1+1) and (2+1) Jets: Cross Sections to \porder{\alpha_s}}
\label{new3}
\label{newnew3}
In this section the calculation of the (1+1) jet
cross section in NLO is reviewed. This is done for two reasons:
these results are needed to calculate jet rates, and their calculation
serves as an illustration of the more complicated case of the NLO
corrections to (2+1) jet production.
As a byproduct, one gets the results for the (2+1) jet Born terms.
The total cross section to \porder{\alpha_s}
for photon exchange has been calculated in \cite{4},
and the (1+1) jet cross section to this order
for all neutral and charged
current processes can be found in \cite{9}. Here the
special case
of the (1+1) jet cross section for the exchange of a transverse photon
is discussed
in detail and results for a longitudinally polarised photon are stated.
It is safe to focus on the QCD corrections for the transverse polarization
of the virtual photon,
since the cross section for longitudinally
polarized photons contributes only about 20\% to the Born term
cross section (see Section \ref{nr}),
and this is expected to be true for the
relative contribution of the longitudinal terms
to the NLO corrections as well.

\noindent
The cross section for eP-scattering differential in $\XB$ and $y$
is given by
\begin{eqnarray}
\label{xsect1}
   \frac{d\sigma_{\!\SS H}}{d\XB dy} & = & \sum_i \int_{\XB}^1 \frac{d\xi}{\xi}
        f_i(\xi) \frac{(4\pi)^\epsilon\,(\SH \XB)^{-\epsilon}\,
        (y(1-y))^{-\epsilon}\,\mu^{4\epsilon}}
        {\Gamma(1-\epsilon)}
        \alpha^2\,\frac{1}{2}\frac{1}{\SH \XB}
        \nonumber\\
   & &  \cdot\,d\mbox{PS}_{\mbox{\small parton}}^{(n)}\,
        \frac{1}{e^2 \mu^{2\epsilon} (2\pi)^{2d}}
        \frac{1}{Q^2} \frac{1}{N'}
        \int d\Omega' \, l^{\mu\nu} \, H_{\mu\nu},
\end{eqnarray}
where $d=(4-2\epsilon)$ is the space-time dimension ($\epsilon\neq 0$
regularises the ultraviolet and infrared divergences
\cite{19,20,21}),
$\mu$ is a mass scale for making the coupling constants dimensionless
in d dimensions, $d\mbox{PS}_{\mbox{\small parton}}^{(n)}$ is the n-parton
phase space
\begin{equation}
d\mbox{PS}_{\mbox{\small parton}}^{(n)}
=(2\pi)^d \prod_{i=1}^n
\frac{d^dp_i\delta(p_i^2)}{(2\pi)^{d-1}}
\delta(p_0+q-\sum_{i=1}^n p_i),
\end{equation}
$\Omega'$ is the volume element of
$d-3$ angles specifying the direction of the outgoing lepton
relative to the outgoing partons, and
\begin{equation}
N' = \int d\Omega'
\end{equation}
is a normalisation constant. The incoming parton carries a fraction $\xi$
of the proton momentum, the $f_i(\xi)$ are the bare parton densities
for partons of flavour $i$,
\begin{equation}
l^{\mu\nu}=k^\mu k'^\nu+k^\nu k'^\mu-k k'g^{\mu\nu}
\end{equation}
is the lepton tensor and $H_{\mu\nu}$ is the hadron tensor (including coupling
constants, colour factors, etc.).

The integration over $\Omega'$ can be
performed. It can be shown by a direct evaluation of the integrals in
$d$-dimensional space that
\begin{eqnarray}
   \frac{1}{Q^2} \frac{1}{N'}
   \int d\Omega' \, l^{\mu\nu} & = &
   \frac{1+(1-y)^2-\epsilon y^2}{2(1-\epsilon)y^2}
   (-g^{\mu\nu}-\epsilon_q^{\mu\nu})\nonumber\\
   & & \!\!\!\!\!\!+
   \frac{4(1-\epsilon)(1-y)+1+(1-y)^2-\epsilon y^2}{2(1-\epsilon)y^2}
   (\epsilon_R^{\mu\nu}+\epsilon_0^{\mu\nu}),
\end{eqnarray}
where
\begin{eqnarray}
   \epsilon_q^{\mu\nu}&=&\frac{1}{Q^2}q^\mu q^\nu,\nonumber\\
   \epsilon_R^{\mu\nu}&=&\frac{1}{Q^2}(q^\mu q^\nu
      +2x_p\big(p_0^\mu q^\nu+q^\mu p_0^\nu)\big),\nonumber\\
   \epsilon_0^{\mu\nu}&=&\frac{4x_p^2}{Q^2}p_0^\mu p_0^\nu,
\end{eqnarray}
and $x_p:=\XB/\xi$.
Because of current conservation $q^\mu H_{\mu\nu}=0$ one has
\begin{equation}
   \epsilon_q^{\alpha\mu} H_{\mu\nu}=\epsilon_R^{\alpha\mu} H_{\mu\nu}=0.
\end{equation}
With the definition $\mbox{tr}H:=h_g:=\left(-g^{\mu\nu}\right)H_{\mu\nu}$,
          $h_0:=\epsilon_0^{\mu\nu}H_{\mu\nu}$
one obtains
\begin{eqnarray}
& &   \frac{1}{Q^2} \frac{1}{N'}
   \int d\Omega' \, l^{\mu\nu} \, H_{\mu\nu}
   \nonu
& &
   =\frac{1+(1-y)^2-\epsilon y^2}{2(1-\epsilon)y^2}h_g
    +\frac{4(1-\epsilon)(1-y)+1+(1-y)^2-\epsilon y^2}{2(1-\epsilon)y^2}
   h_0.
\end{eqnarray}
By defining
\begin{eqnarray}
   \sigma_\lambda & = & \sum_i \int_{\XB}^1 \frac{d\xi}{\xi}
        f_i(\xi) \frac{(4\pi)^\epsilon\,(\SH \XB)^{-\epsilon}\,
        (y(1-y))^{-\epsilon}\,\mu^{4\epsilon}}
        {\Gamma(1-\epsilon)}
        \alpha^2\,\frac{1}{2}\frac{1}{\SH \XB}
        \nonumber\\
   & &  \cdot\,d\mbox{PS}_{\mbox{\small parton}}^{(n)}\,
        \frac{1}{e^2 \mu^{2\epsilon} (2\pi)^{2d}}
        \;h_\lambda
\end{eqnarray}
with $\lambda\in\{g,0\}$ one arrives at
\begin{eqnarray}
   \frac{d\sigma_{\!\SS H}}{d\XB dy} & = &
   \frac{1+(1-y)^2-\epsilon y^2}{2(1-\epsilon)y^2}\,\sigma_g
   \nonumber\\
   & & \!\!\!\!\!\!+
   \frac{4(1-\epsilon)(1-y)+1+(1-y)^2-\epsilon y^2}{2(1-\epsilon)y^2}
   \,\sigma_0.
\end{eqnarray}
In the literature cross sections $\sigma_U,\;\sigma_L$ for unpolarized and
longitudinally polarized photons have been defined. They are related to the
definitions used in this paper
by $\sigma_g=2(1-\epsilon)\sigma_U-\sigma_L,\;\sigma_0=\sigma_L$.

Now the special case of (1+1) jet production is considered.
Fig.~\ref{j1} depicts the Feynman
diagram to \porder{\alpha_s^0}. The diagram for
the 1-loop virtual
correction is shown in fig.~\ref{j2}.
%
%
%
The result for the sum of
these diagrams is
well known \cite{4}:
\begin{eqnarray}
\label{as1b}
   \frac{d\sigma_{\!\SS H}^{\mbox{\small Born\&virt.}}}{d\XB dy} & = &
        \sum_i \int_{\XB}^1 \frac{dx_p}{x_p}
        f_i^{\mbox{ren}}\left(\frac{\XB}{x_p},M_f^2\right)
        \frac{(4\pi)^\epsilon\,(\SH \XB)^{-\epsilon}\,
        (y(1-y))^{-\epsilon}\,\mu^{4\epsilon}}
        {\Gamma(1-\epsilon)}
        \alpha^2\,\frac{1}{2}\frac{1}{\SH \XB}
        \nonumber\\
   & &  \!\!\!\!\!\!\!\!\!\!\!\!\!
        \cdot \frac{1+(1-y)^2-\epsilon y^2}{2(1-\epsilon)y^2}\,
        2\pi\:4(1-\epsilon)\:Q_i^2
\nonu
   & &  \!\!\!\!\!\!\!\!\!\!\!\!\!
        \cdot\bigg\{
        \delta_{qi}
        \bigg[
        1+\left(\frac{4\pi\mu^2}{Q^2}\right)^\epsilon
        \frac{\Gamma(1-\epsilon)}{\Gamma(1-2\epsilon)}
        \frac{\alpha_s}{2\pi} C_F
        \left(-\frac{2}{\epsilon^2}-\frac{3}{\epsilon}\right)
        +\frac{\alpha_s}{2\pi} C_F \left(-8-2\,\zeta(2)\right)
        \bigg]\delta(1-x_p)
\nonu
   & &  \!\!\!\!\!\!\!\!\!\!\!\!\!
        \quad+\left(\frac{4\pi\mu^2}{M_f^2}\right)^\epsilon
        \frac{\Gamma(1-\epsilon)}{\Gamma(1-2\epsilon)}
        \frac{\alpha_s}{2\pi}
        \frac{1}{\epsilon}P_{q\leftarrow i}(x_p)
        \bigg\}
        +\porder{\epsilon}.
\end{eqnarray}
The integration over $\xi$ is rewritten in terms of the
variable $x_p=\XB/\xi$, and the symbol $\delta_{qi}$ restricts the summation
to quark initiated terms.
$\zeta(2)=\pi^2/6$, and $Q_i$ is the charge of the quark
with flavour $i$ normalised to $e$. $\mu$ is the renormalisation scale and
$M_f$ is the factorization scale.
The pole terms in $\epsilon$ proportional to $\delta(1-x_p)$
are infrared singularities that will cancel against infrared
singularities in
the real corrections, and the term proportional to the Altarelli-Parisi
splitting function $P_{q\leftarrow i}(x_p)$ will cancel against
collinear singularities in the real corrections. Note that
the bare parton densities $f_i(\xi)$ are already expressed in terms of
the renormalised parton
densities \cite{4,22} in the $\overline{\mbox{MS}}$-scheme
\begin{equation}
\label{pdredef}
f_i^{\mbox{ren}}\left(\xi,M_f^2\right)=\int_\xi^1\frac{du}{u}
    \Big[\delta_{ij}\delta(1-u)+\frac{\alpha_s}{2\pi}
    \left(-\frac{1}{\epsilon}\right)P_{i\leftarrow j}(u)
    \frac{\Gamma(1-\epsilon)}{\Gamma(1-2\epsilon)}
    \left(\frac{4\pi\mu^2}{M_f^2}\right)^\epsilon
    \Big]f_j(\frac{\xi}{u}).
\end{equation}
The Altarelli-Parisi kernels are
\begin{eqnarray}
P_{q \leftarrow q}(u) & = & C_F\left[
           \frac{1+u^2}{(1-u)_+}+\frac{3}{2}\delta(1-u)
                               \right],
\nonu
P_{g \leftarrow q}(u) & = & C_F\;
           \frac{1+(1-u)^2}{u},
\nonu
P_{g \leftarrow g}(u) & = & 2N_C\left[
           \frac{1}{(1-u)_+}+\frac{1}{u}+u(1-u)-2
                               \right]
                               +\left(
           \frac{11}{6}N_C-\frac{1}{3}N_f
                                \right)\delta(1-u),
\nonu
P_{q \leftarrow g}(u) & = & \frac{1}{2}\left[
           u^2+(1-u)^2\right].
\end{eqnarray}
$N_f$ is the number of quark flavours.

Now the real NLO corrections to (1+1) jet production will be calculated.
The Born terms of \porder{\alpha_s} have to be integrated over the phase space
region that, by the jet definition, is considered to be a (1+1) jet region.
In a first step, suitable variables are defined, then the
(1+1) jet region is determined and finally the integration is performed.
The phase space for the production of 2~partons is constructed in the following
way.
As usual, the
variable
\begin{equation}
z=\frac{P p_1}{P q},
\end{equation}
is defined, where $p_1$ is one of the outgoing partons.
By defining $t=s_{12}/W^2$ ($p_2$ is the momentum of the second outgoing
parton) and $a=\XB+(1-\XB)t$ one obtains
\begin{eqnarray}
\label{c316}
   \int d\mbox{PS}^{(2)} & = & \int d\mbox{PS}^{*(2)}\xi\delta(\xi-a),
\nonu
       \int d\mbox{PS}^{*(2)}
    & = &    \int\frac{(4\pi)^\epsilon}{\Gamma(1-\epsilon)}
        \left(W^2\right)^{-\epsilon}t^{-\epsilon}
        \left(z(1-z)\right)^{-\epsilon}\frac{1}{8\pi}
        \frac{1-\XB}{a} \,dz\,dt,
\end{eqnarray}
where the fact that $\xi=a$ if $p_0=\xi P$ is being used.
By means of the factor $\xi\delta(\xi-a)$ the integral
$\int d\xi/\xi\,f(\xi)$ in the cross section formula \reff{xsect1}
can be performed trivially. The ranges of integration are $z\in{[}0,1]$ and
$t\in{[}0,1]$. The invariants $s_{ij}=2p_i p_j$
expressed in terms of $z$ and $t$ are
\begin{eqnarray}
s_{01} & = & \SH y \big(\XB+(1-\XB)t\big)z,\nonumber\\
s_{02} & = & \SH y \big(\XB+(1-\XB)t\big)(1-z),\nonumber\\
s_{12} & = & \SH y (1-\XB)t.
\end{eqnarray}
The momentum of the target remnant is $p_r=(1-\xi)P$. The {\em observable}
momenta are $p_1$, $p_2$ and $p_r$. Invariants for these momenta are
defined
by $u_{ij}=2p_ip_j/W^2$. In terms of the variables $z$ and $t$ they read
\begin{eqnarray}
u_{r1} & = & (1-t)z,\nonumber\\
u_{r2} & = & (1-t)(1-z),\nonumber\\
u_{12} & = & t.
\end{eqnarray}

Let us define the (2+1) jet region by the condition
$s_{ij}>cW^2$, with $i,j\in\{1,2,r\}$. It is easy to see that
in the $(t,z)$-plane where the allowed phase space is the square
$[0,1]\times [0,1]$ the (2+1) jet region is a triangle given by the following
conditions:
\begin{description}
\item[{\rm (a)}] $\;c < t < 1-2c$,
\item[{\rm (b)}] $\;z_c(t) < z < 1-z_c(t),\;
\mbox{where} \; z_c(t)=\frac{\displaystyle c}{\displaystyle 1-t}$.
\end{description}
The region within the square surrounding this triangle is therefore the
(1+1) jet region.
All regions where the cross section becomes singular are within the (1+1) jet
region thus allowing the factorization of the collinear singularities and their
absorption into the renormalised parton densities.
It should be noted
that for (2+1) jet production the minimal momentum fraction $\xi$
of the parton densities that
can be probed is $\xi_{\mbox{\small min}}(\XB)=\XB+(1-\XB)c$
because of the cut condition on
$s_{12}$. The minimum is $\xi_{\mbox{\small min}}=c$
for $\XB\rightarrow 0$ and
is therefore of order $0.01$.
If one wants to use the gluon initiated (2+1) jet events to
determine gluon densities
at small $x$ a different scale for the jet definition could be
chosen. If the scale is $Q^2$ or
$\left(W^\alpha Q^\beta \sqrt{\SH y}^{1-\alpha-\beta}\right)^2$,
the minimal $\xi$ depending on $\XB$
turns out to be $\xi_{\mbox{\small min}}(\XB)=\XB(1+c)$ and
$\xi_{\mbox{\small min}}(\XB)=\XB+(1-\XB)^\alpha \XB^\beta c$, respectively,
and for sufficiently small $\XB$ and $c$ both give small \
$\xi_{\mbox{\small min}}$. If $\XB$ is fixed, these
definitions are of course equivalent
to the choice of a much smaller cut $c$ in the jet definition scheme based
on $W^2$, and it is therefore important to check whether the
(2+1) jets are still calculated in a regime where perturbation theory is still
valid, i.e., whether the (2+1) jet rate is small enough compared to the
total cross section.

After this short digression the results for
the Born cross sections for the production of 2~partons are stated.
The relevant diagrams are depicted in
fig.~\ref{diag1}
with the graph G replaced by the diagrams of fig.~\ref{diag2}.
%
%
%
Of course, in fig.~\ref{diag1} one would have an additional diagram from an
incoming antiquark, but since the resulting expressions are (except for
the charges) the same as those for incoming quarks,
these diagrams are omitted in the sequel.

\noindent
The traces of the hadron tensor for
quark and gluon initiated processes
are
\begin{eqnarray}
\label{borntot}
   \mbox{tr}H_{\mbox{\small Born, q inc.}}
         & = & L_1 Q_j^2 8 (1-\epsilon) C_F T_q,
   \nonumber\\
   \mbox{tr}H_{\mbox{\small Born, g inc.}}
         & = & L_1 Q_j^2 8 (1-\epsilon) \frac{1}{2} T_g
\end{eqnarray}
with
\begin{eqnarray}
\label{Born1}
        T_q & = &
            (1-\epsilon)
            \left(\frac{s_{02}}{s_{12}}+\frac{s_{12}}{s_{02}}
            \right)
            +\frac{2Q^2s_{01}}{s_{02}s_{12}}+2\epsilon,
        \nonumber\\
        T_g & = &
            \frac{1}{1-\epsilon}
            \left\{
            (1-\epsilon)
            \left(\frac{s_{01}}{s_{02}}+\frac{s_{02}}{s_{01}}
            \right)
            -\frac{2Q^2s_{12}}{s_{01} s_{02}}-2\epsilon
            \right\}
\end{eqnarray}
and
\begin{equation}
L_1=(2\pi)^{2d} g^2 e^2 \mu^{4\epsilon}.
\end{equation}
$p_0$ is always the momentum of the incoming parton, and $p_1$
is the momentum of the outgoing quark.

These formulae are already averaged over the colour degree of freedom
of the
incoming partons.
The factor of $1/(1-\epsilon)$ in the expression for $T_g$ comes from
the fact that a gluon has $2(1-\epsilon)$ helicity states in
$d=(4-2\epsilon)$ dimensions compared to only $2$~helicity
states in $4$~dimensions.

Now the integration of  the cross sections over the (1+1) jet region of the
2-parton phase space can be done
by subtracting the (2+1) jet cross
section from the total cross section to \porder{\alpha_s}. The technical reason
is that this subtraction makes the explicit universal
factorization of the singular
parts in the form of a product of an Altarelli-Parisi splitting function
and the Born term more transparent than the direct integration over
the (1+1) jet region.

\noindent
The total cross section from the real corrections
for transverse photons and quark initiated
processes is \cite{4}:
\begin{eqnarray}
\label{as1q}
   \frac{d\sigma_{\!\SS H}^{\mbox{\small tot.,t,q}}}{d\XB dy} & = &
        \sum_i \int_{\XB}^1 \frac{dx_p}{x_p}
        f_i^{\mbox{ren}}\left(\frac{\XB}{x_p},M_f^2\right)
        \frac{(4\pi)^\epsilon\,(\SH \XB)^{-\epsilon}\,
        (y(1-y))^{-\epsilon}\,\mu^{4\epsilon}}
        {\Gamma(1-\epsilon)}
        \alpha^2\,\frac{1}{2}\frac{1}{\SH \XB}
        \nonumber\\
   & &  \!\!\!\!\!\!\!\!\!\!
        \cdot \frac{1+(1-y)^2-\epsilon y^2}{2(1-\epsilon)y^2}\,
        2\pi\:4(1-\epsilon)\:Q_i^2
        \:\left(\frac{4\pi\mu^2}{Q^2}\right)^\epsilon
        \:\frac{\Gamma(1-\epsilon)}{\Gamma(1-2\epsilon)}
        \:\frac{\alpha_s}{2\pi} \:C_F \:\delta_{qi}
\nonu
   & &  \!\!\!\!\!\!\!\!\!\!
        \cdot\bigg\{
        \frac{2}{\epsilon^2}\delta(1-x_p)
        +\frac{1}{\epsilon}
        \left[
           -\frac{1}{C_F}P_{q\leftarrow q}+3\delta(1-x_p)
        \right]
\nonu
   & &
        +2 \left(\frac{\ln(1-x_p)}{1-x_p}\right)_{+}
        -(1+x_p)\ln(1-x_p)-\frac{3}{2}\frac{1}{(1-x_p)_+}
        -\frac{1+x_p^2}{1-x_p}\ln x_p
\nonu & &
        +3-x_p+\frac{7}{2}\delta(1-x_p)
        \bigg\}
        +\porder{\epsilon}.
\end{eqnarray}
The ``$+$''--prescriptions are defined in \cite{23}
(they are implicitly used for functions defined on the interval $[0,1]$)
and are the result of the subtraction in the collinear regime.
The bare parton densities are expressed in terms of the
renormalised ones and terms of \porder{\alpha_s^2} are dropped.
The corresponding expression for the gluon initiated process is\footnote{
Compared to \cite{4}
there is a difference in the finite parts. It
results from the average over the helicities of an initial gluon.
Here it is assumed that the gluon has $d-2=2(1-\epsilon)$ polarization
states instead of $2$.}
\begin{eqnarray}
\label{as1g}
   \frac{d\sigma_{\!\SS H}^{\mbox{\small tot.,t,g}}}{d\XB dy} & = &
        \sum_{i=1}^{2N_f} \int_{\XB}^1 \frac{dx_p}{x_p}
        f_g^{\mbox{ren}}\left(\frac{\XB}{x_p},M_f^2\right)
        \frac{(4\pi)^\epsilon\,(\SH \XB)^{-\epsilon}\,
        (y(1-y))^{-\epsilon}\,\mu^{4\epsilon}}
        {\Gamma(1-\epsilon)}
        \alpha^2\,\frac{1}{2}\frac{1}{\SH \XB}
        \nonumber\\
   & &  \!\!\!\!\!\!\!\!\!\!
        \cdot \frac{1+(1-y)^2-\epsilon y^2}{2(1-\epsilon)y^2}\,
        2\pi\:4(1-\epsilon)\:Q_i^2
        \:\left(\frac{4\pi\mu^2}{Q^2}\right)^\epsilon
        \:\frac{\Gamma(1-\epsilon)}{\Gamma(1-2\epsilon)}
        \:\frac{\alpha_s}{2\pi} \:\frac{1}{2}\:\frac{1}{1-\epsilon}
\nonu
   & &  \!\!\!\!\!\!\!\!\!\!
        \cdot\bigg\{
        -2\:\frac{1}{\epsilon}
        \:P_{q\leftarrow g}(x_p)
        +\left((1-x_p)^2+x_p^2\right)\ln\frac{1-x_p}{x_p}\bigg\}
        +\porder{\epsilon}.
\end{eqnarray}
Note that the sum over quark flavours reflects the different flavours
that are produced in this
process.\footnote{To make the cancellation of the
collinear divergence transparent, a ``double counting'' is included which
is cancelled by a factor of $1/2$.}

In the sum of
(\ref{as1b}),
(\ref{as1q}) and
(\ref{as1g}) the infrared and collinear singularities cancel, and one is left
with a finite total cross section to \porder{\alpha_s}.
To obtain the (1+1) jet cross section one has to subtract
the (2+1) jet cross section from the total cross section.
Since there are no singularities in the (2+1) jet region,
$\epsilon$ can be set to $0$.
The integration of (\ref{borntot}) over the (2+1) jet region
is not difficult. If all contributions are summed up
one arrives at the finite (1+1) jet cross section. The quark initiated part
is given by
\begin{eqnarray}
   \frac{d\sigma_{\!\SS H}^{\mbox{\small (1+1),t,q}}}{d\XB dy} & = &
        \sum_{i=1}^{2N_f} \int_{\XB}^1 \frac{dx_p}{x_p}
        \:Q_i^2\:f_i^{\mbox{ren}}\left(\frac{\XB}{x_p},M_f^2\right)
        \alpha^2\,\frac{1}{2}\frac{1}{\SH \XB}
        \frac{1+(1-y)^2}{2y^2}\,
        \cdot 2\pi\cdot 4
\nonu
   & &  \!\!\!\!\!\!\!\!\!\!
        \cdot\bigg\{1+C_F\:\frac{\alpha_s}{2\pi}
        \bigg[\left(-8-2\zeta(2)\right)\delta(1-x_p)
\nonu
   & &
        +2 \left(\frac{\ln(1-x_p)}{1-x_p}\right)_{+}
        -(1+x_p)\ln(1-x_p)-\frac{3}{2}\frac{1}{(1-x_p)_+}
        -\frac{1+x_p^2}{1-x_p}\ln x_p
\nonu & &
        +3-x_p+\frac{7}{2}\delta(1-x_p)
        +\frac{1}{C_F}\:\ln\frac{Q^2}{M_f^2}P_{q\leftarrow q}(x_p)
\nonu & &
        -\bigg(\left[
          \frac{1}{2}\frac{1}{1-x_p}-2\frac{x_p}{1-x_p}
        \right]\left(1-2z_c(t(x_p))\right)
\nonu & & \quad \quad
        +\left[
          1-x_p+2\frac{x_p}{1-x_p}
        \right]\ln\frac{1-z_c(t(x_p))}{z_c(t(x_p))}
        \bigg)\Xi_{c\le t(x_p)\le 1-2c}
        \bigg]\bigg\},
\end{eqnarray}
and the gluon initiated processes contribute
\begin{eqnarray}
   \frac{d\sigma_{\!\SS H}^{\mbox{\small (1+1),t,g}}}{d\XB dy} & = &
        \sum_{i=1}^{2N_f} \int_{\XB}^1 \frac{dx_p}{x_p}
        \:Q_i^2\:f_g^{\mbox{ren}}\left(\frac{\XB}{x_p},M_f^2\right)
        \alpha^2\,\frac{1}{2}\frac{1}{\SH \XB}
        \frac{1+(1-y)^2}{2y^2}\,
        \cdot 2\pi\cdot 4
\nonu
   & &  \!\!\!\!\!\!\!\!\!\!
        \cdot\:\frac{1}{2}\:\frac{\alpha_s}{2\pi}
        \bigg[
        \left((1-x_p)^2+x_p^2\right)
        \left(\ln\frac{1-x_p}{x_p}-1\right)
        +2\:\ln\frac{Q^2}{M_f^2}P_{q\leftarrow g}(x_p)
\nonu & &
        -\bigg(\!\!
        -\left(1-z_c(t(x_p))\right)
         +\left(1-2x_p(1-x_p)\right)
\nonu & & \quad \quad \quad \quad \quad \quad \quad \quad \quad \quad \cdot
        \ln\frac{1-z_c(t(x_p))}{z_c(t(x_p))}
        \bigg)\Xi_{c\le t(x_p)\le 1-2c}
        \bigg],
\end{eqnarray}
where $\Xi_A$ is the characteristic function of the set specified in $A$
restricting the integration to that set and
$t(x_p)$ is the variable $t$ introduced above given by
\begin{equation}
t(x_p)=\frac{\XB}{1-\XB}\frac{1-x_p}{x_p}.
\end{equation}
The logarithms depending on the factorization scale
that cancel part of the scale dependence of the parton densities
are indicated explicitly.
There are no such terms depending on the renormalisation scale because
the Born term is of \porder{\alpha_s^0}.

\noindent
The corresponding terms for a virtual photon with longitudinal polarization
arise at \porder{\alpha_s}. They are given by
\begin{eqnarray}
   \frac{d\sigma_{\!\SS H}^{\mbox{\small (1+1),l,q}}}{d\XB dy} & = &
        \sum_{i=1}^{2N_f} \int_{\XB}^1 \frac{dx_p}{x_p}
        \:Q_i^2\:f_i^{\mbox{ren}}\left(\frac{\XB}{x_p},M_f^2\right)
        \alpha^2\,\frac{1}{2}\frac{1}{\SH \XB}
        \frac{4(1-y)+1+(1-y)^2}{2y^2}\,
        \cdot 2\pi\cdot 4
\nonu
   & &  \!\!\!\!\!\!\!\!\!\!
        \cdot C_F\:\frac{\alpha_s}{2\pi}
        \bigg[
        x_p-\bigg(x_p\left(1-2z_c(t(x_p))\right)
        \bigg)\Xi_{c\le t(x_p)\le 1-2c}
        \bigg]
\end{eqnarray}
and
\begin{eqnarray}
   \frac{d\sigma_{\!\SS H}^{\mbox{\small (1+1),l,g}}}{d\XB dy} & = &
        \sum_{i=1}^{2N_f} \int_{\XB}^1 \frac{dx_p}{x_p}
        \:Q_i^2\:f_g^{\mbox{ren}}\left(\frac{\XB}{x_p},M_f^2\right)
        \alpha^2\,\frac{1}{2}\frac{1}{\SH \XB}
        \frac{4(1-y)+1+(1-y)^2}{2y^2}\,
        \cdot 2\pi\cdot 4
\nonu
   & &  \!\!\!\!\!\!\!\!\!\!
        \cdot\:\frac{1}{2}\:\frac{\alpha_s}{2\pi}
        \bigg[
        2x_p(1-x_p)-\bigg(
        2x_p(1-x_p)\left(1-2z_c(t(x_p))\right)
        \bigg)\Xi_{c\le t(x_p)\le 1-2c}
        \bigg]
\end{eqnarray}
for incoming quarks and gluons, respectively.

Now the jet recombination scheme ambiguity
which arises from the problem of the mapping of a
``jet phase space'' of massive jets in NLO onto a phase space of
massless partons in leading order is discussed.
The real corrections for the (1+1) jet cross section have been obtained by an
integration of the Born terms for the production of 2 partons over some
region of phase space specified by the jet cut definition.
To cancel the infrared and collinear
singularities, one has to define effective (1+1) jet variables
that allow the identification of the corresponding
singularities in the virtual corrections and in the contribution
from the redefinition of the parton densities. This procedure is a map
of the (1+1) jet region of the 2-parton phase space onto the
(1+1) jet phase space. In principle this map is ambiguous, but fortunately
the the difference between different maps is 0 for vanishing jet cut.
Here the following scheme
is tacitly assumed: If a cluster algorithm is
applied to a 2-parton event that looks like a (1+1) jet event, then
the final result of this clustering will be two clusters
with momenta $p_A$ and $p_B$, one of them
being massless and one of them being massive.
The (1+1) jet Born term and the virtual correction always result in
two massless clusters $p_C$ and $p_D$. The identification is
done by mapping $(p_A+p_B)^2$ onto $(p_C+p_D)^2$. In this special case
both expressions are equal to $W^2$. This quantity can be determined from
electron variables alone, so what is essentially done is the identification
of processes in which the electrons have the same momenta. This, however,
may correspond to very different situations on the parton level,
since e.g. a radiated final state gluon can be collinear to the remnant jet
resulting in a remnant with a large momentum or can be collinear to
the outgoing quark.
The situation is similar in the case of the corrections to the (2+1) jet
cross section. In that case, however, the identification involves
two variables and not only one, and so the identification using $W^2$ is not
sufficient.

\section{(2+1) Jets: Virtual Corrections}
\label{old3}
In this section the calculation of the
virtual corrections to the production of (2+1) jets is described.
The one-loop corrections to the graphs from fig.~\ref{diag2} are given in
fig.~\ref{diag3}. The one-loop diagrams from fig.~\ref{diag4} contribute
to the wave function renormalisation and are taken into account
in the counter term.
%
%
%
To obtain the \porder{\alpha_s^2} corrections
the diagrams in fig.~\ref{diag3} must be
multiplied by the Born diagrams. The resulting topologies can be
divided into three classes:

\vspace{0.2cm}
\noindent
\makebox[1.0cm][l]{(I)} QED-like graphs with colour factor $N_C C_F^2$,

\vspace{0.2cm}
\noindent
\makebox[1.0cm][l]{(II)} QED-like graphs with colour factor
     $N_C C_F\left(C_F-\frac{1}{2}N_C\right)$,

\vspace{0.2cm}
\noindent
\makebox[1.0cm][l]{(III)} non-abelian graphs with colour factor
     $-\frac{1}{2} N_C^2 C_F$,

\vspace{0.2cm}
\noindent
resulting in terms proportional to the colour factors
$N_C C_F^2$ and $N_C^2 C_F$.
The sum of the virtual \porder{\alpha_s^2} corrections
averaged over the colour degree of freedom of the incoming
parton is

\begin{eqnarray}
\label{virt1}
   \mbox{tr}H_{\mbox{\small v, q inc.}}
        & = & L_1 L_2 Q_j^2 8 (1-\epsilon)
        \big\{C_F^2 E_{\mbox{\small 1,q}}
               -\frac{1}{2}N_C C_F E_{\mbox{\small 2,q}}
        \nonumber\\
        & &    +\left(\frac{1}{\epsilon}
                 +\log\frac{Q^2}{\mu^2}\right)
                 \left(\frac{1}{3}N_f-\frac{11}{6}N_C
                 \right) C_F T_q
         \big\} + \porder{\epsilon},
\end{eqnarray}

\begin{eqnarray}
\label{virt2}
     \mbox{tr}H_{\mbox{\small v, g inc.}}
        & = & L_1 L_2 Q_j^2 8 (1-\epsilon)
        \big\{\frac{1}{2} C_F E_{\mbox{\small 1,g}}
               -\frac{1}{4}N_C E_{\mbox{\small 2,g}}
        \nonumber\\
        & &    +\left(\frac{1}{\epsilon}
                 +\log\frac{Q^2}{\mu^2}\right)
                 \left(\frac{1}{3}N_f-\frac{11}{6}N_C
                 \right) \frac{1}{2} T_g
         \big\} + \porder{\epsilon},
\end{eqnarray}
where
\begin{equation}
L_2=\frac{\alpha_s}{2\pi}
    \left(\frac{4\pi\mu^2}{-q^2-i\eta}
    \right)^\epsilon
    \frac{\Gamma(1-\epsilon)}{\Gamma(1-2\epsilon)}.
\end{equation}
Here $N_f$ is the number of active quark flavours in the fermion loops
and $\mu^2$ is the renormalisation scale (it is understood that the running
$\alpha_s$ is always evaluated at the scale $\mu^2$).
The contributions $E_{\mbox{\small 1,q}}$,
$E_{\mbox{\small 2,q}}$, $E_{\mbox{\small 1,g}}$ and
$E_{\mbox{\small 2,g}}$ come from
the calculation of the matrix elements of the graphs in fig.~\ref{diag3}.
The explicit expressions for the
$E_{\mbox{\small i,q}}$ and $E_{\mbox{\small i,g}}$ are collected in
Appendix \ref{appb}.
The trace calculations of the matrix elements were done with the
help of REDUCE \cite{24}. Then the loop integrals
were performed by an insertion of one-loop
tensor structure integrals (see appendix \ref{appa}). Here one has to be
careful with respect to the imaginary parts of Spence functions and
logarithms which are important because $q^2<0$. The results
obtained here have been checked
against those of the \epem-case \cite{25}. This was possible because
all infinitesimal imaginary
parts from the propagators were kept in the formulae.

\begin{sloppypar}
In (\ref{virt1}), (\ref{virt2}) the counter terms (in the
$\overline{\mbox{MS}}$-scheme) to cancel UV singularities
(see \cite{25}) is already added.
Some
$1/\epsilon^2$- and $1/\epsilon$-poles remain.
These divergences are due to the
IR singularities of the loop corrections.
\end{sloppypar}

\noindent
For the processes with an incoming quark the following variables are defined:
\begin{equation}
\label{defz1}
z_q:=\frac{p_1 p_0}{p_0 q}, \quad
z_g:=\frac{p_2 p_0}{p_0 q}, \quad
x_p:=\frac{\XB}{a}.
\end{equation}

\noindent
The divergent parts are
\begin{eqnarray}
   \mbox{tr}H_{\mbox{\small v, q inc.}} & = &
        L_1 L_2 Q_j^2 8 (1-\epsilon) C_F T_q \nonu
       & & \cdot\bigg\{
             C_F\bigg[
             \left(-\frac{1}{\epsilon^2}+\frac{1}{\epsilon}
             \left(\ln\frac{1-z_g}{x_p}-\frac{3}{2}
             \right)\right)\nonu
          & &  + \left(-\frac{1}{\epsilon^2}+\frac{1}{\epsilon}
             \left(\ln\frac{z_q a}{\XB}-\frac{3}{2}
             \right)\right)
          \bigg]\nonumber\\
          & & -\frac{1}{2}N_C\bigg[
             \left(\frac{2}{\epsilon^2}+\frac{1}{\epsilon}
             \left(\ln\frac{x_p^2}{(1-x_p)(1-z_q)}+2
             \right)\right)
            + \frac{1}{\epsilon}\cdot\frac{5}{3}\nonu
          & &  + \frac{1}{\epsilon}\ln\frac{1-z_g}{1-x_p}
            + \frac{1}{\epsilon}\ln\frac{z_q}{1-z_q}
          \bigg]\nonumber\\
          & & + N_f \cdot\frac{1}{3} \cdot\frac{1}{\epsilon}
        \bigg\} + \porder{\epsilon^0}.
\end{eqnarray}

\noindent
For the processes with an incoming gluon the variables are similar:
\begin{equation}
\label{defz2}
z_q:=\frac{p_1 p_0}{p_0 q}, \quad
z_{\overline q}:=\frac{p_2 p_0}{p_0 q}, \quad
x_p:=\frac{\XB}{a}.
\end{equation}

\noindent
The divergent parts are
\begin{eqnarray}
   \mbox{tr}H_{\mbox{\small v, g inc.}} & = &
        L_1 L_2 Q_j^2 8 (1-\epsilon) \frac{1}{2} T_g \nonu
       & & \cdot\bigg\{
             C_F\bigg[
             2\left(-\frac{1}{\epsilon^2}+\frac{1}{\epsilon}
             \left(\ln\frac{1-x_p}{x_p}-\frac{3}{2}
             \right)\right)
          \bigg]\nonumber\\
          & & -\frac{1}{2}N_C\bigg[
             \frac{1}{\epsilon}
             \ln\frac{1-x_p}{1-z_{\overline q}}
             +\frac{1}{\epsilon}
             \ln\frac{1-x_p}{1-z_q}
          \nonu
          & &  +\left(\frac{2}{\epsilon^2}
               + \frac{1}{\epsilon}
                 \left(-\ln\frac{a^2 z_q(1-z_q)}{\XB^2}
            + \frac{11}{3}\right)\right)
          \bigg]\nonumber\\
          & & + N_f \cdot \frac{1}{3} \cdot \frac{1}{\epsilon}
        \bigg\} + \porder{\epsilon^0}.
\end{eqnarray}

\noindent
They will cancel against divergent terms from the real corrections.


\section{(2+1) Jets: Final State Real Corrections}
\label{old4}
\label{sect4}
To $\porder{\alpha_s^2}$ one has to consider the contributions from the Born
terms in fig.~\ref{diag41} integrated over the (2+1) jet phase space region
in the 3-parton phase space.
%

\noindent
Again there are graphs with an incoming quark and an incoming gluon.
The generic diagrams are shown in
fig.~\ref{diag42}. There are, of course, additional contributions
with incoming antiquarks; their structure is identical to the
quark-initiated processes.
%

The integrations become singular if the integrand contains a propagator
whose denominator vanishes in the integration region. The
method of partial
fractions is used to separate initial and final state singularities.
This allows the identification of the terms proportional to
$\porder{c^0}$, $\porder{\ln c}$ and $\porder{\ln^2c}$ (c is the jet cut).
In this section the final state singularities are considered, the initial
state singularities are treated in the next section.

\noindent
For the process
\begin{equation}
   e^-(k)+\mbox{proton}(P) \rightarrow e^-(k')
   + \mbox{target remnant}(p_r)
   + \mbox{parton}(p_1)
   + \mbox{parton}(p_2)
   + \mbox{parton}(p_3)
\end{equation}
a parametrization of the phase space
of the outgoing partons with momenta $p_i$ is needed.
The target remnant
with momentum $p_r=(1-\xi)P$ is described by the variable $\xi$.
The parametrization is chosen such that the integration over the region
$s_{12}<c$ ($p_1$ and $p_2$
being collinear or $p_1$
being soft or $p_2$
being soft) is simple.
In close analogy to calculations in \epem-annihilation it is reasonable
to describe the two particle phase space of $p_1$ and $p_2$ in the
CM frame of these momenta (see fig.~\ref{diag44}).
%

\noindent
Let $p_0$ be the momentum of the incoming parton. One can define a variable
$z$ by
\begin{equation}
   z:=\frac{p_0 p_3}{p_0 q}
\end{equation}
that describes the phase space of $p_3$ (the azimuthal dependence
is contained in the lepton phase space, and the remaining third integration
is trivial because of energy conservation). One can define a
polar angle $\theta$ given
by $\theta:=\angle(\vec{p}_1,\vec{p}_0)$ in the CM frame of $p_1$ and
$p_2$ and an azimuthal angle $\varphi$ by the angle between the planes
spanned by $\vec{p}_0$, $\vec{p}_1$ and $\vec{p}_0$, $\vec{p}_3$,
respectively. Let $\chi$ be defined by $\chi:=\angle(\vec{p}_0,\vec{p}_3)$
and
\begin{eqnarray}
   b & := & \frac{1}{2} (1-\cos\theta)
\\
   \label{ref414}
   d & := & \frac{1}{2} (1-\cos\chi)
\\
   e & := & b + d - 2bd - 2 \sqrt{b(1-b)d(1-d)} \cos\varphi.
\end{eqnarray}

\noindent
With the normalisation factors
\begin{eqnarray}
N_\varphi & := & \frac{\pi 4^\epsilon \Gamma(1-2\epsilon)}
                      {\Gamma^2(1-\epsilon)}
             = \int_0^\pi \sin^{-2\epsilon}\varphi d\varphi
\\
N_b & := & \frac{\Gamma^2(1-\epsilon)}
                      {\Gamma(2-2\epsilon)}
             = \int_0^1 \left(b(1-b)\right)^{-\epsilon} db
\end{eqnarray}
the 3-parton phase space in $d=4-2\epsilon$ dimensions is
\begin{eqnarray}
\label{old418}
\int d\mbox{PS}^{(3)}
  & = & \int \frac{(16\pi^2)^\epsilon}{128\pi^3\Gamma(2-2\epsilon)}
             s_{12}^{-\epsilon} ds_{12} z^{-\epsilon}
             \left(\SH y(\xi-\XB)(1-z)-s_{12}\right)^{-\epsilon}
\nonu & &
        \cdot
        dz \frac{1}{N_\varphi}\sin^{-2\epsilon}\varphi d\varphi
           \frac{1}{N_b}\left(b(1-b)\right)^{-\epsilon} db
\nonu & = & \int d\mbox{PS}^{*(2)} \int d\mbox{PS}^{*(r)}.
\end{eqnarray}
$d\mbox{PS}^{*(2)}$ is defined in eq.~(\ref{c316}), and
\begin{eqnarray}
  d\mbox{PS}^{*(r)} & = & a\delta(\xi-a) L_2 \frac{2\pi}{\alpha_s}
  \mu^{-2\epsilon} \frac{1}{2} \frac{1}{8\pi^2} \frac{Q^2}{x_p}
  x_p^\epsilon \frac{1}{1-2\epsilon} d\mu_F,
\\
  d\mu_F & = & \left(1-\frac{s_{12}}{\SH y (\xi-\XB)(1-z)}\right)^{-\epsilon}
         r_{12}^{-\epsilon} dr_{12} \frac{1}{N_\varphi}
         \sin^{-2\epsilon}\varphi d\varphi
\nonu & &
         \cdot\frac{1}{N_b}\left(b(1-b)\right)^{-\epsilon} db,
\\
  r_{ij} & := & \frac{s_{ij}}{\SH y \xi}.
\end{eqnarray}
$d\mbox{PS}^{*(2)}$ contains the variables $z$ and $t=s_{12}/W^2$
that are identified with
the corresponding (2+1) jet variables. The phase space for 3 particles
factorizes as a product of a phase space for 2 particles $d\mbox{PS}^{*(r)}$
and an effective
phase space for a particle and a cluster $d\mbox{PS}^{*(2)}$.
The invariants $r_{ij}$ can be expressed in the variables
$t$, $z$, $b$, $x_p=\XB/\xi$, $\varphi$ and $r_{12}$:
\begin{eqnarray}
   r_{01} & = & (1-z) b,
\nonu
   r_{02} & = & (1-z) (1-b),
\nonu
   r_{03} & = & z,
\nonu
   r_{13} & = & (1-x_p-r_{12}) e,
\nonu
   r_{23} & = & (1-x_p-r_{12}) (1-e).
\end{eqnarray}
The variable $d$ in eq.~(\ref{ref414}) is given by
\begin{equation}
   d = \frac{z}{1-z} \frac{r_{12}}{1-x_p-r_{12}}.
\end{equation}
The phase space boundaries are
\begin{eqnarray}
& & \XB \in {[}0,1{]}, \quad \xi \in {[}\XB,1{]}, \quad t \in {[}0,1{]},
\nonu
& & \varphi \in {[}0,\pi{]}, \quad b \in {[}0,1{]},
        \quad r_{12} \in {[}0,(1-x_p)(1-z){]}.
\end{eqnarray}

The products of the diagrams
in fig.~\ref{diag41}
with the complex conjugated diagrams can be classified in classes A, B, ..., H
with different colour factors (see tab.~\ref{tab41}).
An explicit calculation shows that the colour classes G and H are regular
when integrated over the 2-particle phase space
$d\mbox{PS}^{*(r)}$ and therefore vanish for $c\rightarrow 0$. Therefore
these classes are not considered here.

\begin{table}[htb]
\hspace{1.5cm}
\begin{tabular}{|c|c|c|c|c|c|c|c|} \hline
                & I & II & III & IV & V & VI & VII \\ \hline
\rule[-1mm]{0mm}{6mm}$\overline{\mbox{I}}$
& A & *  & *   &    &   &    &     \\ \hline
\rule[-1mm]{0mm}{6mm}$\overline{\mbox{II}}$
& B & A  & *   &    &   &    &     \\ \hline
\rule[-1mm]{0mm}{6mm}$\overline{\mbox{III}}$
& C & E  & D   &    &   &    &     \\ \hline
\rule[-1mm]{0mm}{6mm}$\overline{\mbox{IV}}$
&   &    &     & F  & * & *  & *   \\ \hline
\rule[-1mm]{0mm}{6mm}$\overline{\mbox{V}}$
&   &    &     & G  & F & *  & *   \\ \hline
\rule[-1mm]{0mm}{6mm}$\overline{\mbox{VI}}$
&   &    &     & G  & H & F  & *   \\ \hline
\rule[-1mm]{0mm}{6mm}$\overline{\mbox{VII}}$
&   &    &     & H  & G & G  & F   \\ \hline
\end{tabular}
\hspace{1.5cm}
\begin{tabular}{|c|c|}\hline
Class & Colour Factor \\ \hline
A & $C_F^2 N_C$ \\ \hline
B & $N_C C_F (C_F-N_C/2)$ \\ \hline
C & $-1/2 N_C^2 C_F$ \\ \hline
D & $N_C^2 C_F$ \\ \hline
E & $1/2 N_C^2 C_F$ \\ \hline
F & $1/2 N_C C_F$ \\ \hline
G & $N_C C_F (C_F-N_C/2)$ \\ \hline
H & $1/2 N_C C_F$ \\ \hline
\end{tabular}
\caption{\label{tab41} Colour classes and colour factors
                       of the real corrections.}
\end{table}

\noindent
The calculation of the spin sum for external gluons has been
performed with the formula
\begin{equation}
\sum_{\lambda=0}^{d-1} \epsilon_\mu^\lambda \epsilon_\nu^\lambda=-g_{\mu\nu}.
\end{equation}
To cancel the contributions from scalar and longitudinal gluons
one has to subtract diagrams with external ghost lines. The longitudinal
and scalar contributions then drop out because of the Slavnov-Taylor
identities (see \cite{26}).

The matrix elements have been calculated with REDUCE in $d=4-2\epsilon$
dimensions. In principle they could be obtained from the results
in \epem-scattering \cite{27,28,29}. However, the results of the
procedure to obtain partial fractions are
different here since some of the
invariants pick up a sign because of the crossing prescriptions.
Here only the results of the factorization of the IR divergent
terms
\begin{equation}
   {\cal M}_{\mbox{singular}}=K\cdot T_{q/g}.
\end{equation}
are stated.
$T_{q/g}$ is the Born term (\ref{Born1})
with incoming quark and gluon, respectively,
and $K$ is a singular kernel whose integration is divergent in $d=4$
dimensions. In $d=4-2\epsilon$ dimensions the result of the
phase space integral is of the form
$a/\epsilon^2 +B/\epsilon+C$. $A$ and $B$ do not depend on
the invariant mass cut $c$. C contains terms of the form
$\ln c$ and $\ln^2c$ which diverge for $c\rightarrow 0$.
The contributions from the final state singularities
are divided into seven classes (see tab.~\ref{tab43}).
The results for the traces of the hadronic tensor
$\tr H_{F_i}$ for the seven colour classes are given explicitly
in appendix \ref{kernelfss}.

\begin{table}[htb]
\begin{center}
\begin{tabular}{|c|c|c|c|}\hline
Class & incoming parton & product of diagrams & colour factor \\ \hline
$F_1$ & quark & I$\cdot$I, II$\cdot$II, II$\cdot$I & $N_C C_F^2/N_C$ \\ \hline
$F_2$ & quark & II$\cdot$I, III$\cdot$I, III$\cdot$II &
              $(-1/2) N_C^2 C_F /N_C$ \\ \hline
$F_3$ & quark & III$\cdot$I, III$\cdot$II & $(-1/2) N_C^2 C_F /N_C$ \\ \hline
$F_4$ & quark & III$\cdot$III & $N_C^2 C_F /N_C$ \\ \hline
$F_5$ & quark & IV$\cdot$IV, V$\cdot$V, VI$\cdot$VI, VII$\cdot$VII &
              $(1/2) N_C C_F /N_C$ \\ \hline
$F_6$ & gluon & I$\cdot$I, II$\cdot$II, II$\cdot$I &
              $N_C C_F^2/(2N_C C_F)$ \\ \hline
$F_7$ & gluon & II$\cdot$I, III$\cdot$I, III$\cdot$II &
              $(-1/2) N_C^2 C_F /(2N_C C_F)$ \\ \hline
\end{tabular}
\caption{\label{tab43} Colour factors.}
\end{center}
\end{table}

The singular kernels are integrated over the (2+1) jet like region in
phase space. The condition for partons j, k to form a jet is
$s_{jk}\le cW^2$. This condition can be rephrased in terms of
the variables
\begin{equation}
r_{jk} := \frac{s_{jk}}{\SH y \xi}
\end{equation}
as
\begin{equation}
r_{jk}\le c\frac{1-\XB}{\XB}x_p.
\end{equation}
The phase space boundary is given by
$r_{jk}\le(1-x_p)(1-z)$.
So the (2+1) jet region is specified by
\begin{equation}
r_{jk}\le\alpha:=\min \Big\{(1-x_p)(1-z),c\frac{1-\XB}{\XB}x_p\Big\}, \quad
b \in {[}0,1{]}, \quad\varphi\in{[}0,\pi{]}.
\end{equation}
The phase space integrals are listed in appendix \ref{appc}.
One obtains
\begin{equation}
\int_{\mbox{(2+1) jet}}d\mbox{PS}^{*(r)}\tr H_{F_i}=a\delta(\xi-a)
                       L_1 L_2 8(1-\epsilon) F_i.
\end{equation}
for the corrections from the final state singularities.
The explicit expressions are collected in appendix \ref{appd}.

\section{(2+1) Jets: Initial State Real Corrections}
\label{old5}
In this section the contributions from the initial
state singularities are described.
In the case of the final state singularities the
momentum fraction $\xi$ of the incoming parton is a fixed parameter
in the matrix elements. In the case of the initial
state singularities, however, there is the additional problem
that $\xi$ is an integration variable and that $\xi$ is an argument
of the parton densities $f_i(\xi)$.
Although there is no expression for $f_i(\xi)$ in a closed form,
the problem can be solved \cite{4}
by a Taylor expansion around the singular
point and the well known ``+''-prescriptions \cite{23}
\begin{equation}
D_+(g):=\int_0^1du D(u) \left(g(u)-g(1)\right).
\end{equation}

A similar parametrization of the phase space as
in the case of the final state singularities is used, but the
effective (2+1) jet variables are defined in a different way.
Let $p_0$ be the momentum of the incoming parton
and $p_3$ the momentum of an outgoing parton. For the calculation
of the singularity resulting from $s_{03}\rightarrow 0$ $p_1$ and
$p_2$ are the momenta of the partons to be identified with the effective
(2+1) jet momenta.
Variables $z'$, $z$ and $t$ are defined by
\begin{equation}
z' := \frac{p_0 p_3}{p_0 q}, \quad z:=\frac{p_0 p_1}{p_0 q},
\quad t:= \frac{s_{12}}{W^2}.
\end{equation}
With the conventions from Section \ref{sect4} one obtains
\begin{equation}
b=\frac{z}{1-z'}.
\end{equation}

\noindent
A factor of unity
\begin{equation}
1=\int d\xi'\delta(\xi-\xi')
\end{equation}
is inserted in eq.~(\ref{old418})
to display a change of variables
\begin{equation}
\sigma:=\frac{1-\frac{t}{1-z'}-\frac{1-\xi'}{1-\XB}}
             {1-\frac{t}{1-z'}}
\end{equation}
explicitly.
As a result one obtains
\begin{eqnarray}
\int d\mbox{PS}^{(3)} & = & \int d\mbox{PS}^{*(2)}\int d\mbox{PS}^{*(r)},
\\
\int d\mbox{PS}^{*(r)} & = & \delta(\xi-\xi') L_2 \frac{2\pi}{\alpha_s}
                             \mu^{-2\epsilon}\frac{1}{2}
                             \frac{1}{8\pi^2}W^2 H(z') d\mu_I,
\\
H(z') & = & (1-z')^{-2+2\epsilon}\left(1-\frac{z'}{1-t}\right)^{1-\epsilon}
            \left(1-\frac{z'}{1-z}\right)^{-\epsilon} = 1+\porder{z'},
\\
d\mu_I & = & \frac{\Gamma(1-2\epsilon)}{\Gamma^2(1-\epsilon)}
             a \left(\frac{1-\XB}{\XB} \right)^{-\epsilon}
             (1-t)^{1-\epsilon} \sigma^{-\epsilon}
             d\sigma z'^{-\epsilon}dz' \frac{1}{N_\varphi}
             \sin^{-2\epsilon}\varphi d\varphi.
\end{eqnarray}
The $\delta$-function $\delta(\xi-\xi')$ can be used to perform
the integration over the parton densities. $d\mu_I$ is the measure for
the singular integrations.
The invariants $t_{ij}:=s_{ij}/W^2$ can be expressed in terms of the
phase space variables:
\begin{eqnarray}
t_{01} & = & (\nu-\zeta)z,
\nonu
t_{02} & = & (\nu-\zeta)(1-z-z'),
\nonu
t_{03} & = & (\nu-\zeta)z',
\nonu
t_{12} & = & t,
\nonu
t_{13} & = & (1-\zeta-t)e,
\nonu
t_{23} & = & (1-\zeta-t)(1-e),
\end{eqnarray}
where
\begin{eqnarray}
\nu & = & \frac{1}{1-\XB}, \quad
          \zeta=(1-\sigma)\left(1-\frac{t}{1-z'} \right),
\\
e & = & b + d + 2bd -2\sqrt{b(1-b)d(1-d)}\cos\varphi,
\\
d & = & \frac{z'}{1-z'}\frac{r}{1-\frac{\XB}{\xi'}-r}.
\end{eqnarray}
It should be noted that
\begin{equation}
\xi=a+(1-a)\sigma + \porder{z'}.
\end{equation}
The phase space boundaries are given by
\begin{eqnarray}
\XB \in {[}0,1{]}, \quad \xi, \xi' \in {[}\XB,1{]}, \quad t \in {[}0,1{]},
\quad z \in {[}0,1{]},
\nonu
\varphi \in {[}0,\pi{]}, \quad \sigma \in {[}0,1{]}, \quad
        0\le z' \le \min\big\{1-z,1-t\big\}.
\end{eqnarray}

\noindent
The factorization of the divergent parts is performed in the form
\begin{equation}
   {\cal M}_{\mbox{singular}}=K\cdot T_{q/g}.
\end{equation}
up to terms that vanish for $c \rightarrow 0$ after the integration.
Therefore one can, for example, set $H(z')$ identically to $1$.

\noindent
The contributions from the initial state singularities
are divided into seven classes (see tab.~\ref{tab51}).
In the case of the initial state singularities the incoming parton of the
(3+1) jet graph is not necessarily the incoming parton of the
(factorized) (2+1) jet process. This is the familiar fact that
the quark parton densities modify
the evolution of the gluon density and {\em vice versa}.
In tab.~\ref{tab51}
a list of the incoming partons of both processes is added.
The explicit expressions of the singular kernels
$\tr H_{I_i}$ are collected in appendix \ref{kerneliss}.

\begin{table}[htb]
\begin{center}
\begin{tabular}{|c|c|c|c|c|}\hline
Class & (3+1)
      & (2+1) & product of diagrams & colour factor \\ \hline
$I_1$ & quark & quark
              & I$\cdot$I, II$\cdot$II, II$\cdot$I & $N_C C_F^2/N_C$ \\ \hline
$I_2$ & quark & quark & II$\cdot$I, III$\cdot$I, III$\cdot$II &
              $(-1/2) N_C^2 C_F /N_C$ \\ \hline
$I_3$ & quark & gluon & IV$\cdot$IV, V$\cdot$V, VI$\cdot$VI, VII$\cdot$VII &
              $ (1/2) N_C C_F /N_C$ \\ \hline
$I_4$ & gluon & quark & I$\cdot$I, II$\cdot$II, II$\cdot$I &
              $N_C C_F^2/(2N_C C_F)$ \\ \hline
$I_5$ & gluon & quark & II$\cdot$I, III$\cdot$I, III$\cdot$II &
              $(-1/2) N_C^2 C_F /(2N_C C_F)$ \\ \hline
$I_6$ & gluon & gluon & II$\cdot$I, III$\cdot$I, III$\cdot$II &
              $(-1/2) N_C^2 C_F /(2N_C C_F)$ \\ \hline
$I_7$ & gluon & gluon & III$\cdot$III &
              $ N_C^2 C_F /(2N_C C_F)$ \\ \hline
\end{tabular}
\caption{\label{tab51} Colour factors.}
\end{center}
\end{table}

The boundaries of the (2+1) jet phase space region are given
by the invariant mass cut condition.
For the initial state singularities, the variable $z'$ is the crucial
variable that determines the singularity structure. The matrix elements
become singular for $z'=0$. Let ``$r$'' be the label for the target remnant
jet. The invariant $t_{rj}$ is given by
$t_{rj}=\left((1-\xi)/\xi\right) t_{0j}$.
If $p_r$ and $p_3$ are the momenta combined into a jet, then the invariants
of the effective (2+1) jet event are $t_{12}$, $t_{1r3}$, $t_{2r3}$.
The cut conditions read
\begin{eqnarray}
   t_{r3} & = & \frac{1-\xi}{1-\XB}z' \le c,
\\
   t_{12} & = & t \ge c,
\\
   t_{1r3} & = & t_{13}+t_{1r}+t_{3r} = (1-\zeta-t)e
                +\frac{1-\xi}{1-\XB}(z+z') \ge c,
\\
   t_{2r3} & = & t_{23}+t_{2r}+t_{3r} = (1-\zeta-t)(1-e)
                +\frac{1-\xi}{1-\XB}(1-z) \ge c.
\end{eqnarray}
Especially the last two of these conditions are too complicated to
be used in an analytical calculation since they involve a restriction
of the azimuthal angle integration. Therefore the contributions
from the initial state singularities are integrated
up to the phase space boundary
(so $\varphi\in{[}0,\pi{]}, z'\in{[}0,\min\{1-z,1-t\}{]}$) by keeping
the effective (2+1) jet variables $z$ and $t$ fixed.
The (3+1) jet contribution is then subtracted after a numerical integration.
Since the integral including the parton densities cannot be performed
analytically, this is not a serious restriction.
The shape of the phase space regions in the $(z',\sigma)$--plane is given in
fig.~\ref{diag53}.
%

The poles and double poles in $\epsilon$ characterising IR ($z'=\sigma=0$)
and collinear ($z'=0$) singularities can be calculated in the integration
over the full $(z',\sigma)$-plane.
The phase space integrals are given in appendix \ref{appe}.
In the formulae given there the upper limit of the
$z'$-integration is $\beta=\min\{1-z,1-t\}$.
In the integrals $\sigma$ is used as an integration variable.
The integral involving the parton densities is of the form
\begin{equation}
\label{eqn533}
   \int_0^1d\sigma \frac{f(\xi(\sigma,z',\ldots),Q^2)}
                        {\xi(\sigma,z',\ldots)} D(\sigma).
\end{equation}
Here $D$ is a generalized function depending on $\sigma$ and the other
jet variables. $f(\xi(\sigma,z',\ldots),Q^2)$ is expanded in a
Taylor series in $z'$ and all terms of order \porder{z'} that
do not contribute in the approximation used here
are neglected. One obtains
\begin{equation}
f(\xi(\sigma,z',\ldots)=f(a+(1-a)\sigma,Q^2)+\porder{z'}.
\end{equation}
$a$ is the momentum fraction of the incoming parton of the
factorized Born term. With the definition
\begin{equation}
u:=\frac{a}{\xi}=\frac{a}{a+(1-a)\sigma}+\porder{z'}
\end{equation}
eq.~(\ref{eqn533}) can be rewritten as
\begin{equation}
\int_0^1\frac{du}{u}f(\frac{a}{u},Q^2) D\left(\sigma(u)\right)\frac{1}{1-a}.
\end{equation}
Since $D$ is a generalized function, one has to take care
for the boundary terms of the variable transformation $\sigma\rightarrow u$.

\noindent
Finally one obtains
for the real corrections from the initial state singularities
\begin{equation}
   \int_{\mbox{(2+1) jet}\;\cup\;
           \mbox{(3+1) jet}}d\mbox{PS}^{*(r)} \tr H_{I_i}
         = \int_a^1 \frac{du}{u}\xi \delta\left(\xi-\frac{a}{u}\right)
           L_1 L_2 8(1-\epsilon) I_i.
\end{equation}
The explicit expressions for the $I_i$ are given in appendix \ref{appf}.

\section{(2+1) Jets: Finite Cross Sections}
\label{old6}
\label{newnew7}
In the preceeding sections the calculation of the
Born terms \porder{\alpha_s}, the virtual corrections \porder{\alpha_s^2}
and the real corrections \porder{\alpha_s^2} has been described.
In the sum of the virtual
and real corrections the IR singularities cancel, and the remaining collinear
singularities are absorbed into the parton densities by the redefinition
eq.~(\ref{pdredef}).
The final cross sections is then free of divergencies.
The partonic cross sections must be multiplied with
the different flavour factors of the 14 classes of diagrams
and integrated over the momentum fraction of the incoming parton.
Let the charge of the quark of flavour $i$ be $q_i=Q_i e$, where
$i=1$ stands for d-quarks, $i=2$ for $\overline{\mbox{d}}$, $i=3$ for
u-quarks, and so on. Let $f_i(\xi,M_f^2)$ be the parton density
of flavour $i$, $f_g(\xi,M_f^2)$ the gluon density and $N_f$ the number
of flavours.
Then the flavour factors are
\begin{eqnarray}
\mbox{\makebox[5.3cm][l]{$
H_{F_1},H_{F_2},H_{F_3},H_{F_4},H_{I_1},H_{I_2}:$}}
& & \sum_{i=1}^{2N_f}Q_i^2 f_i(\xi,M_f^2),
\nonu
\mbox{\makebox[5.3cm][l]{$
H_{F_5}:$}} & & N_f\sum_{i=1}^{2N_f}Q_i^2 f_i(\xi,M_f^2),
\nonu
\mbox{\makebox[5.3cm][l]{$
H_{F_6},H_{F_7},H_{I_4},H_{I_5},H_{I_6},H_{I_7}:$}}
& & \sum_{i=1}^{N_f}Q_{2i-1}^2 f_g(\xi,M_f^2),
\nonu
\mbox{\makebox[5.3cm][l]{$
H_{I_3}:$}} & & \sum_{i=1}^{N_f}Q_{2i-1}^2\sum_{j=1}^{2N_f}f_j(\xi,M_f^2).
\end{eqnarray}
Here the factors $Q_j^2$ are included which are already present
in the terms $H_{F_i}$ and $H_{I_i}$.

\noindent
The Born terms and virtual corrections with incoming quarks
are multiplied by
\begin{equation}
\sum_{i=1}^{2N_f}Q_i^2 f_i(\xi,M_f^2),
\end{equation}
those with an incoming gluon by
\begin{equation}
\sum_{i=1}^{N_f}Q_{2i-1}^2 f_g(\xi,M_f^2).
\end{equation}

\section{Numerical Results}
\label{old7}
\label{new8}
\label{nr}
In this section the results of the numerical evaluation
of the jet cross sections are presented.
The finite (1+1) jet cross section has been
calculated in
Section \ref{newnew3},
the (2+1) jet cross section in Section \ref{newnew7}.

The matrix elements for (1+1) and (2+1) jet production
are implemented in the program
PROJET 3.3 \cite{30} which uses
the multidimensional adaptive integration routine
VEGAS \cite{31,32}
for the numerical integrations. The parton density parametrizations
are from the package PAKPDF \cite{33}.
PROJET 3.3 allows the integration over bins in
$\XB$, $y$, $W^2$ and $Q^2$. Furthermore, acceptance cuts on the angles
of the outgoing lepton and the outgoing jets in the
laboratory frame can be applied.

To be definite, the following parameters are used.
In the case of HERA the CM energy is $E_{\mbox{\small CM}}=295\;\GeV$,
whereas the CM energy of the E665 experiment is
$E_{\mbox{\small CM}}=31\;\GeV$.
In the latter case the lepton phase space is restricted by
2~\GeV~$<$~$Q$~$<$~5~\GeV{},
20~\GeV~$<$~$W$~$<$~40~\GeV{}
and $0.05\;<y\;<0.95$.
Unless otherwise stated,
the renormalisation and the factorization scales are set to
$\mu=M_f=Q$,
and the parton densities are from the MRS, set D-- parametrization
\cite{34},
which is presently favoured by structure function measurements
at HERA.
The number of flavours in the final state is set to 5.

The dependence of the jet cross sections
on the jet cut $c$ is shown in fig.~\ref{fig12}. The jet
definition scheme is $s_{ij}\grkl cW^2$
($s_{ij} \grkl M^2$ means that two clusters have to be combined if their
invariant mass is smaller than $M^2$ and have to be considered as two
separate clusters if their invariant mass is larger than $M^2$).
The phase space of the outgoing lepton for HERA is assumed to be
$0.001\,<\,\XB\,<\,1$,
10~\GeV~$<$~$W$~$<$~295~\GeV, and
3.16~\GeV~$<$~$Q$~$<$~10~\GeV{} (a),
10~\GeV~$<$~$Q$~$<$~31.6~\GeV{} (b),
31.6~\GeV~$<$~$Q$~$<~$100~\GeV{} (c).
The graph in fig.~\ref{fig12} (d) is for E665.
The (1+1) jet
Born cross section does not depend on the jet cut, because the
condition $s_{r1}>cW^2$ is always trivially satisfied for $c<1$. The (1+1) jet
cross section in NLO decreases with decreasing cut $c$, because the total cross
section to \porder{\alpha_s} is independent of $c$ while the (2+1) jet cross
section to \porder{\alpha_s} strongly increases with decreasing cut.
The (2+1) jet cross section in NLO is comparable to the (2+1) jet cross section
on the Born level as long as the jet cut is not too small.
At $c\,\approx\,0.007$ the NLO (2+1) jet cross section starts to decrease, and
will go to $-\infty$ for $c\rightarrow 0$ because of dominant terms
$\sim\,-\ln^2c$ in the NLO corrections.
If one thinks of $cW^2$ as a new scale in the cross section,
this behaviour
(an extremum of the cross section at some value of this scale)
can be interpreted as a stabilsation with respect to a change
in this scale.
If the difference between the cross
section on the Born level and the cross section in NLO is too large
this is a sign of the breakdown of fixed order perturbation theory.
However, if
$c\,\ge\,0.01$ this does not seem to be the case, and therefore the region
of values for $c$ for phenomenological studies should start here.
The contribution from the longitudinal polarisation of the
virtual photon is always small (of the order of 20\%) compared to
the transverse cross section. Therefore
using the transverse contributions in NLO and the longitudinal contributions
on the Born level should be accurate (the relative
magnitude of the corrections to the longitudinal cross sections are expected to
be of the same order as in the transverse case, see also \cite{14}).
The cut dependence of the jet cross sections in all the three regions in $Q$
for HERA studied here is similar, up to the absolute normalisation.

The jet rates $R_{\mbox{\small(2+1)}}$ in fig.~\ref{fig13} are defined by
$R_{\mbox{\small(2+1)}}=\sigma_{\mbox{\small(2+1)}}/
\sigma_{\mbox{\small tot}}$.
In the case of Born
terms, $\sigma_{\mbox{\small tot}}$ is given by the total cross section
to \porder{\alpha_s^0}, which is
equal to the (1+1) jet cross section on the Born
level.
In the case of the NLO corrections,
$\sigma_{\mbox{\small tot}}$ is given by the total cross section
to \porder{\alpha_s}, which is equal to the
sum of the (1+1) and (2+1) jet cross sections to \porder{\alpha_s}.
This definition has the advantage that the denominator is always
independent of the jet cut.
$R_{\mbox{\small(2+1), Born}}$ is strongly increasing with decreasing jet cut
$c$. $R_{\mbox{\small(2+1), NLO}}$ is smaller than
$R_{\mbox{\small(2+1), Born}}$
in the bin of smaller values of $Q$ (a). In the bin of larger
values of $Q$ (c),
$R_{\mbox{\small(2+1), NLO}}$ is larger than
$R_{\mbox{\small(2+1), Born}}$ for $c\,>\,0.015$ and smaller for
$c\,<\,0.015$.

Now two different jet definition schemes are compared, fig.~\ref{fig14}.
Here we use the HERA CM energy.
The lepton phase space is given by
3.16~\GeV~$<$~$Q$~$<$~20~\GeV,
10~\GeV~$<$~$W$~$<$~295~\GeV.
In (a) the scheme is $s_{ij}\grkl cW^2$, and in
(b) the cut condition is $s_{ij}\grkl c\left(W^\alpha Q^\beta
\sqrt{\SH y}^{1-\alpha-\beta}\right)^2$
with
$\alpha=0.8$, $\beta=0.2$.
The (2+1) jet cross sections on the Born level
in the scheme (b) are always larger (for the
same value of $c$) because the absolute scale of the cut
in (b) is always smaller.
Furthermore, the NLO starts to deviate considerably from the Born level
at $c\,\approx\,0.02$ in the jet definition scheme (b).
Therefore, if this scheme is used, larger values
of $c$ are advised. If the parameter $\beta$ is too small,
then the NLO cross sections are frequently negative (for small absolute values
of the cut scale, fixed order perturbation theory breaks down, this is similar
to the case of small c in the $cW^2$ scheme, see above).

The results for the scale dependence of the cross sections
at HERA energies are shown in
fig.~\ref{fig15} (a)--(c)
($0.001\,<\,\XB\,<\,1$,
5~\GeV~$<$~$Q$~$<$~100~\GeV,
10~\GeV~$<$~$W$~$<$~295~\GeV, $c\,=\,0.02$).
In principle, the renormalisation scale $\mu$
and the factorization scale $M_f$ are arbitrary.
These scales give rise to logarithms of the form
$\ln(\mu^2/M^2)$ and
$\ln(M_f^2/M^2)$ in the cross section, where $M$ is some mass scale
in the process. The logarithms are potentially large and spoil perturbation
theory if the renormalisation and factorization scales
are not of the same order
of magnitude as $M$. In order to study the behaviour of the cross sections
for a change of the scale, the renormalisation scale
(fig.~\ref{fig15} (a), (c)) and
the factorization scale (fig.~\ref{fig15} (b), (c)) are varied
in the form of $\rho Q$, where $\rho$
is a parameter in the range 0.2 to 5.
If only the renormalisation scale is varied (a), the (1+1) jet cross section
on the Born level is constant, because it is of \porder{\alpha_s^0}.
The (2+1) jet cross section on the Born level
possesses a large scale dependence
of $\pm 40\%$ in the range of $\rho$ given above.
The NLO correction to the (2+1) jet cross section reduces this
scale dependence considerably, because there is a term logarithmic in
$\mu$ that cancels a part of the scale dependence of the running coupling
in the Born term such that the overall dependence on $\mu$ is
(formally) of
$\porder{\alpha_s^3}$.
The (1+1) jet cross section in NLO is scale dependent because of the
running coupling constant, and there is no mechanism (i.e., no
explicit logarithmic term in $\mu$) that would cancel this dependence
(the reason for this is discussed in Section \ref{new3}).
Fortunately, the total cross section to \porder{\alpha_s} is less
scale dependent than the (1+1) jet cross section.
The dependence on the factorization scale is shown in fig.~\ref{fig15} (b).
The (1+1) jet Born term is strongly scale dependent. Because the parton
densities are redefined when the NLO contributions are calculated, there
is a term that makes the dependence on $M_f$ formally of
\porder{\alpha_s^2} in NLO.
A similar cancellation takes place for the (2+1) jet cross section.
If both the renormalisation and the factorization scale are varied
(c), the overall picture is that, compared to the Born level, the NLO
cross sections are less scale dependent.
In fig. \ref{fig16new} (a) the scale dependence of the jet rates
is shown.\footnote{The three curves for the NLO corrected
terms should intersect
at $\rho=1$. The small difference at $\rho=1$ is due to small
fluctuations from the spline fit to the Monte Carlo results.}
It is evident that the NLO results have a much smaller scale dependence
than the results on the Born level.
The same graphs for HERA for the range of $Q$ given by
100~\GeV~$<$~$Q$~$<$~200~\GeV{}
are shown in
fig.~\ref{fig15} (d)--(f),
and the corresponding graph for the jet rates is
fig. \ref{fig16new} (b).
The scale dependent cross sections and jet rates for the
kinematical region of the E665 experiment (for a jet cut $c=0.04$)
are shown in fig.~\ref{fig15} (g)--(i) and
fig. \ref{fig16new} (c).

Finally the dependence on the parton densities at HERA for two
different jet definition schemes (fig.~\ref{fig16}) is discussed,
with parameters
5~\GeV~$<$~$Q$~$<$~295~\GeV,
10~\GeV~$<$~$W$~$<$~295~\GeV, $c\,=\,0.02$.
The parametrizations HMRS set B \cite{35},
MT set B1 \cite{36} and the more recent ones MRS sets D0 and D--
are chosen for comparison.
The two sets of curves in fig.~\ref{fig16} are for the cross sections
$\XB d\sigma/d\XB$ for (1+1) and (2+1) jets in NLO.
The jet definition scheme in
(a) is $s_{ij}\grkl cW^2$.
For values of $\XB$ smaller than 0.01 the different parametrizations clearly
predict different (1+1)
jet cross sections. The (2+1) jet cross section
differential in $\XB$ is insensitive to a variation
of the parametrization. The reason is that in the $cW^2$ scheme
all contributions to the (2+1) jet cross section come from
$\xi>c$, $\xi$ being the momentum fraction of the incoming parton,
as discussed in Section \ref{old2}. Since $c=0.02$, there is
very small variation in the (2+1) jet cross section
because there is not
much difference in the parametrizations for $\xi>0.02$.
The situation is different in the scheme
(b) $s_{ij}\grkl c\left(W^\alpha Q^\beta
\sqrt{\SH y}^{1-\alpha-\beta}\right)^2$
with
$\alpha=0.7$, $\beta=0.3$. Here the (2+1) jet cross section receives
contributions from the parton densities at $\xi\,<\,c$ as well, and
therefore the (2+1) jet cross section depends on the parametrization.
Using such a scheme might therefore be a possibility to measure the
gluon density $f_g(\xi,M_f^2)$ for small $\xi$ via (2+1) jet cross sections
by a a subtraction of the quark initiated contribution
from the total (quark and gluon initiated)
(2+1) jet cross section.

\section{Summary and Conclusions}

In this paper the calculation of (1+1) and (2+1) jet cross
sections in deeply inelastic
electron proton scattering has been described.
The jet definition includes the target remnant jet and is based on a modified
JADE cluster algorithm.
The inclusion of the proton remnant in the jet definition scheme is a
consistent way to define 'exclusive' jet cross section for
the production of (n+1) jets because of the possibility
of collinear emission of partons in the direction of the
target remnant jet.

The cross sections are studied for HERA and E665 energies in detail.
The jet cut dependence suggests that, if $cW^2$ is used
as the mass scale in the invariant
jet definition, the jet cut $c$ should be larger than 0.01 to avoid large
NLO corrections that could invalidate a fixed order perturbative expansion.
In the $cW^2$ scheme, the (2+1) jet cross section depends on the parton
densities $f_i(\xi,M_f^2)$ for $\xi>c$ only, even for very small $\XB$.
If one wishes to probe the parton densities at smaller values of $\xi$,
a different jet definition scheme has to be used.
In the proposed region for $c$, the NLO corrections are at most
of the order of 30\%, and even smaller for very large $Q^2$.

\begin{sloppypar}
A set of jet definition
schemes that could be useful is given by the cut condition
$s_{ij} \grkl c\left(W^\alpha Q^\beta \sqrt{\SH y}^{1-\alpha-\beta}\right)^2$,
where $\alpha$ and $\beta$ are some parameters in the range of $[0,1]$.
By comparing the results for different parametrizations of parton densities
it is explicitly shown that such a jet definition scheme
gives a strong dependence of the (2+1) jet cross section on the chosen
parametrization. However, it must be studied whether such a jet definition is
experimentally feasible and useful for the determination of the gluon density.
\end{sloppypar}

An important point is the scale dependence of the calculated
cross section. It is a
general phenomenon that leading order cross sections that depend on
the strong coupling constant and scale dependent parton densities
are {\em strongly}
scale dependent. This leads to a theoretical uncertainty because,
in principle, the renormalisation and factorization scales are arbitrary
(although they should be chosen to be of the order of some physical
scale in the process) and the variation of the cross section
with respect to changes in the scales can be interpreted as being due to
(unknown) higher order corrections because the cross section to all orders must
be independent of the scales. The NLO
corrections usually improve the situation
because terms arise that cancel part of the scale dependence of the leading
order. This desirable feature is present in the calculation
described here, as has been shown
explicitly by a variation of the scales as multiples of $Q^2$. The scale
dependence is significantly reduced.
It can be concluded that the NLO corrections reduce the
theoretical uncertainties
of the leading order and should provide well defined jet cross sections
that could be useful in experimental analyses.

\section{Acknowledgements}

I have benefitted from interesting disscussions with Ch.~Berger,
Th.~Brodkorb, J.~Conrad, I.~Hinchliffe, G.~Ingelman, J.G.~K\"orner,
G.~Kramer, Z.~Kunszt, N.~Magnussen,
R.~Nisius, C.~Salgado,
H.~Spiesberger, G.~Schuler and
P.~Zerwas.
I would like to thank W. Buchm\"uller (DESY) and
I. Hinchliffe (LBL) for the warm hospitality extended to me at DESY and LBL
where part of this work has been done.
Moreover I
wish to thank I. Hinchliffe and M. Luty for a critical reading
of the manuscript.
M. Klasen and D. Michelsen pointed
out some typographic errors in \cite{12}.
\newpage
\begin{appendix}
\section{Massless 1-Loop Tensor Structure Integrals}
\label{appa}
This appendix contains the results for the tensor structure integrals
\begin{eqnarray}
I_{2\{0,1\mu\}}(r) & := & \int d^dk\frac{\{1,k_\mu\}}
                   {(k^2+i\eta)((k-r)^2+i\eta)},
\nonu
I_{3\{0,1\mu,2\mu\nu\}}(r,p) & := & \int d^dk\frac{\{1,k_\mu,k_\mu k_\nu\}}
                   {(k^2+i\eta)((k-r)^2+i\eta)((k-r-p)^2+i\eta)},
\\
I_{40}(p_1,p_2,p_3) & := & \int d^dk\frac{1}
                   {(k^2+i\eta)((k+p_2)^2+i\eta)((k-p_1)^2+i\eta)
                    ((k-p_1-p_3)^2+i\eta)}
\nonumber
\end{eqnarray}
needed for the evaluation of the virtual corrections for graphs with
massless particles. It is assumed that $p^2=p_1^2=p_2^2=p_3^2=0$, but
not necessarily $r^2=0$.
The integrals are regularized by dimensional regularisation ($d=4-2\epsilon$)
to take care of UV and IR divergences.
$i\eta$ is an infinitesimal imaginary part, $\eta>0$. It is
convenient to define the functions
\begin{eqnarray}
F(r,p) & := & \left(\frac{r^2+2pr+i\eta}{q^2+i\eta}\right)^{-\epsilon}
             -\left(\frac{r^2+i\eta}{q^2+i\eta}\right)^{-\epsilon},
\\
G(r,p) & := & \frac{-q^2}{2pr}\left[
              \left(\frac{r^2+2pr+i\eta}{q^2+i\eta}\right)^{1-\epsilon}
             -\left(\frac{r^2+i\eta}{q^2+i\eta}\right)^{1-\epsilon}
              \right],
\\
H(r,p) & := & \left(\frac{-q^2}{2pr}\right)^2\left[
              \left(\frac{r^2+2pr+i\eta}{q^2+i\eta}\right)^{2-\epsilon}
             -\left(\frac{r^2+i\eta}{q^2+i\eta}\right)^{2-\epsilon}
              \right].
\end{eqnarray}
The calculation with Feynman parametrizations gives (compare \cite{25})
\begin{eqnarray}
I_{20}(r) & = & \frac{i\pi^{2-\epsilon}\Gamma(1+\epsilon)\Gamma^2(1-\epsilon)}
                     {\epsilon^2\Gamma(1-2\epsilon)}
                (-q^2-i\eta)^{-\epsilon}(\epsilon+2\epsilon^2)
                \left(\frac{r^2+i\eta}{q^2+i\eta}\right)^{-\epsilon}
                +\porder{\epsilon},
\\
I_{21\mu}(r) & = & \frac{i\pi^{2-\epsilon}\Gamma(1+\epsilon)
                                          \Gamma^2(1-\epsilon)}
                     {\epsilon^2\Gamma(1-2\epsilon)}
                (-q^2-i\eta)^{-\epsilon}(\frac{\epsilon}{2}+\epsilon^2)
                \left(\frac{r^2+i\eta}{q^2+i\eta}\right)^{-\epsilon}
                r_\mu
                +\porder{\epsilon},
\\
I_{30}(r,p) & = & \frac{i\pi^{2-\epsilon}\Gamma(1+\epsilon)
                                         \Gamma^2(1-\epsilon)}
                     {\epsilon^2\Gamma(1-2\epsilon)}
                (-q^2-i\eta)^{-\epsilon}
                \frac{1}{2pr}F(r,p)
                +\porder{\epsilon},
\\
I_{31\mu}(r,p) & = & \frac{i\pi^{2-\epsilon}\Gamma(1+\epsilon)
                                          \Gamma^2(1-\epsilon)}
                     {\epsilon^2\Gamma(1-2\epsilon)}
                (-q^2-i\eta)^{-\epsilon}
                \cdot\bigg\{
                (1+\epsilon+2\epsilon^2)\frac{1}{2pr}F(r,p)r_\mu
\nonu
           & &  +(1+\epsilon+2\epsilon^2)\frac{1}{2pr}
                \left(-\frac{r^2}{2pr}F(r,p)+(\epsilon+\epsilon^2)
                                             G(r,p)\right)p_\mu
                \bigg\}
                +\porder{\epsilon},
\\
I_{32\mu\nu}(r,p) & = & \frac{i\pi^{2-\epsilon}\Gamma(1+\epsilon)
                                          \Gamma^2(1-\epsilon)}
                     {\epsilon^2\Gamma(1-2\epsilon)}
                (-q^2-i\eta)^{-\epsilon}
                \left(1+\frac{3}{2}\epsilon+3\epsilon^2\right)\frac{1}{2pr}
\nonu
           & &  \cdot\bigg\{
                -\frac{1}{4}\left(\epsilon+\frac{3}{2}\epsilon^2\right)
                \,2pr\,G(r,p) g_{\mu\nu}
\nonu
           & &  +F(r,p) r_\mu r_\nu
\nonu
           & &  +\left[
                   \left(\frac{r^2}{2pr}\right)^2F(r,p)
                   -2(\epsilon+\epsilon^2)\frac{r^2}{2pr}G(r,p)
                   -\frac{1}{2}\left(\epsilon+\frac{1}{2}\epsilon^2\right)
                   H(r,p)
                 \right]p_\mu p_\nu
\nonu
           & &  +\left[
                   -\frac{r^2}{2pr}F(r,p)
                   +(\epsilon+\epsilon^2)G(r,p)
                 \right](p_\mu r_\nu + p_\nu r_\mu)
                \bigg\}
                +\porder{\epsilon},
\\
I_{40}(p_1,p_2,p_3) & = &
                \frac{i\pi^{2-\epsilon}\Gamma(1+\epsilon)\Gamma^2(1-\epsilon)}
                     {\epsilon^2\Gamma(1-2\epsilon)}
                (-q^2-i\eta)^{-\epsilon}
                \frac{2}{(q^2)^2y_{12} y_{13}}
\nonu
           & &  \cdot\bigg\{
                   1-\epsilon\left(\lln(y_{12})+\lln(y_{13})\right)
\nonu & &
                   +\epsilon^2
                   \left(
                   \frac{1}{2}\lln^2(y_{12})+\frac{1}{2}\lln^2(y_{13})
                   +R(y_{12},y_{13})
                   \right)
                \bigg\}
                +\porder{\epsilon}.
\end{eqnarray}
In $I_{40}$ the momentum $q$ is defined by $q=p_1+p_2+p_3$, and the invariants
$y_{ij}$ are given by
$y_{ij}:=2p_ip_j/q^2$.
The function $R$ is given by
\begin{equation}
   R(x,y)=\lln(x)\;\lln(y)-\lln(x)\;\lln(1-x)-\lln(y)\;\lln(1-y)
         -\spence(x)-\spence(y)+\zeta(2).
\end{equation}
$\lln(x)$ is the natural logarithm with an additional prescription for
arguments on the cut ${[}-\infty,0{]}$
\begin{equation}
\lln(x):=\lim_{\eta\searrow 0}\ln(x+\mbox{sgn}(q^2)\mbox{sgn}(1-x) i\eta),
\end{equation}
and $\spence$ is defined by
\begin{equation}
\spence := \lim_{\eta\searrow 0}\dilog(x+\mbox{sgn}(q^2)\mbox{sgn}(1-x) i\eta),
\end{equation}
where
$\dilog$ is the complex dilogarithm
\begin{equation}
   \dilog(z) = - \int_0^z du \frac{\ln(1-u)}{u}.
\end{equation}

It can easily be seen where the $i\eta$-prescription is important.
For $q^2>0$ it fixes the imaginary part of the factor
$(-q^2-i\eta)^{-\epsilon}$. Expanded up to \porder{\epsilon^2}
this gives the well known $\pi^2$-terms (in combination with
$1/\epsilon^2$-poles) in \epem-annihilation and in the Drell-Yan
process.
In deeply inelastic scattering $(-q^2-i\eta)^{-\epsilon}$ has no imaginary
part. However, the functions $F$, $G$ and $H$ give rise to $\pi^2$-terms
in combination with poles $1/\epsilon, 1/\epsilon^2$.

\section{Virtual Corrections}
\label{appb}

In this appendix the explicit expressions for the virtual corrections
are given.
For the processes with an incoming quark the following invariants
are defined:
\begin{eqnarray}
\yqi & = & \frac{z_q}{x_p}=\frac{1-z_g}{x_p},\nonu
\yig & = & \frac{1-z_q}{x_p}=\frac{z_g}{x_p},\nonu
\yqg & = & \frac{1-x_p}{x_p},
\end{eqnarray}
where the variables $z_q$ and $z_g$ are defined in eq.~(\ref{defz1})
and $x_p=\XB/\xi$.
One then obtains
\begin{eqnarray}
& & E_{\mbox{1,q}}=\left[-\frac{2}{\epsilon^2}+\frac{1}{\epsilon}
                     \left(2\ln\yqi-3\right)\right]\cdot T_q
\nonu
& & + \left(-2\zeta(2)-\ln^2\yqi-8\right)\cdot T_q
\nonu
& & +4\ln(\yqi)\left(\frac{2\yqi}{-\yqg+\yig}
                    +\frac{\yqi^2}{(-\yqg+\yig)^2}
               \right)
\nonu
& & +\ln(\yqg)\left(\frac{4\yqi-2\yqg}{\yqi+\yig}
                  +\frac{\yqg\yig}{(\yqi+\yig)^2}
              \right)
\nonu
& & +\ln(\yig)\left(\frac{4\yqi+2\yig}{\yqi-\yqg}
                  +\frac{\yqg\yig}{(\yqi-\yqg)^2}
              \right)
\nonu
& & +2\frac{\yqi^2+(\yqi+\yig)^2}{\yqg\yig} R'(\yqi,-\yqg)
    +2\frac{\yqi^2+(\yqi-\yqg)^2}{\yqg\yig} R'(\yqi,\yig)
\nonu
& & +\yqi \left(
         \frac{4}{-\yqg+\yig}+\frac{1}{\yqi-\yqg}+\frac{1}{\yqi+\yig}
         \right)
\nonu
& & + \frac{\yqi}{\yqg}
    - \frac{\yqi}{\yig}
    + \frac{\yqg}{\yig}
    + \frac{\yig}{\yqg},
\end{eqnarray}

\begin{eqnarray}
& & E_{\mbox{2,q}}=\left[\frac{2}{\epsilon^2}+\frac{1}{\epsilon}
                     \left(2\ln\yqi-2\ln\yqg-2\ln\yig\right)\right]\cdot T_q
\nonu
& & + \left(2\zeta(2)-\ln^2\yqi
            +\left(\ln^2\yqg-\pi^2\right)+\ln^2\yig
            +2R'(-\yqg,\yig)\right)\cdot T_q
\nonu
& & -\Bigg[\ln(\yqg)\frac{2\yqg}{\yqi+\yig}
    +\ln(\yig)\frac{-2\yig}{\yqi-\yqg}
\nonu
& & \quad +4\ln(\yqi)\left(\frac{\yqi^2}{(-\yqg+\yig)^2}
                  +\frac{2\yqi}{-\yqg+\yig}
              \right)
\nonu
& & \quad -2\left(
    -  \frac{\yqg}{\yig}
    -  \frac{\yig}{\yqg}
    -  \frac{\yqi}{\yqg}
    +  \frac{\yqi}{\yig}
    -  \frac{2\yqi}{-\yqg+\yig}
    \right)
\nonu
& & \quad +2R'(\yqi,-\yqg)\frac{\yqi^2+(\yqi+\yig)^2}{\yqg\yig}
\nonu
& & \quad +2R'(\yqi,\yig) \frac{\yqi^2+(\yqi-\yqg)^2}{\yqg\yig}
    \Bigg].
\end{eqnarray}

For the processes with an incoming gluon the following variables are defined:
\begin{eqnarray}
\yqi  & = & \frac{z_q}{x_p}=\frac{1-z_{\overline q}}{x_p},\nonu
\yqbi & = & \frac{z_{\overline q}}{x_p}=\frac{1-z_q}{x_p},\nonu
\yqqb & = & \frac{1-x_p}{x_p}.
\end{eqnarray}
The variables $z_q$ and $z_{\overline q}$ are defined in eq.~(\ref{defz2}).
For these processes one obtains
\begin{eqnarray}
& & E_{\mbox{1,g}}=\left[-\frac{2}{\epsilon^2}+\frac{1}{\epsilon}
                     \left(2\ln\yqqb-3\right)\right]\cdot T_g
\nonu
& & + \left(-2\zeta(2)-(\ln^2\yqqb-\pi^2)-8\right)\cdot T_g
\nonu
& & +4\ln(\yqqb)\left(\frac{-2\yqqb}{\yqi+\yqbi}
                    +\frac{\yqqb^2}{(\yqi+\yqbi)^2}
               \right)
\nonu
& & +\ln(\yqi)\left(\frac{-4\yqqb+2\yqi}{-\yqqb+\yqbi}
                  -\frac{\yqi\yqbi}{(-\yqqb+\yqbi)^2}
              \right)
\nonu
& & +\ln(\yqbi)\left(\frac{-4\yqqb+2\yqbi}{-\yqqb+\yqi}
                  -\frac{\yqi\yqbi}{(-\yqqb+\yqi)^2}
              \right)
\nonu
& & -2\frac{\yqqb^2+(-\yqqb+\yqbi)^2}{\yqi\yqbi} R'(-\yqqb,\yqi)
    -2\frac{\yqqb^2+(-\yqqb+\yqi)^2}{\yqi\yqbi} R'(-\yqqb,\yqbi)
\nonu
& & -\yqqb \left(
         \frac{4}{\yqi+\yqbi}+\frac{1}{-\yqqb+\yqi}+\frac{1}{-\yqqb+\yqbi}
         \right)
\nonu
& & + \frac{\yqqb}{\yqi}
    + \frac{\yqqb}{\yqbi}
    - \frac{\yqi}{\yqbi}
    - \frac{\yqbi}{\yqi},
\end{eqnarray}

\begin{eqnarray}
& & E_{\mbox{2,g}}=\left[\frac{2}{\epsilon^2}+\frac{1}{\epsilon}
                     \left(2\ln\yqqb-2\ln\yqi-2\ln\yqbi\right)\right]\cdot T_g
\nonu
& & + \left(2\zeta(2)
            -\left(\ln^2\yqqb-\pi^2\right)+\ln^2\yqi+\ln^2\yqbi
            +2R'(\yqi,\yqbi)\right)\cdot T_g
\nonu
& & +\ln(\yqi)\frac{-2\yqi}{-\yqqb+\yqbi}
    +\ln(\yqbi)\frac{-2\yqbi}{-\yqqb+\yqi}
\nonu
& & +4\ln(\yqqb)\left(\frac{\yqqb^2}{(\yqi+\yqbi)^2}
                  -\frac{2\yqqb}{\yqi+\yqbi}
              \right)
\nonu
& & -2\left(
       \frac{\yqi}{\yqbi}
    +  \frac{\yqbi}{\yqi}
    -  \frac{\yqqb}{\yqi}
    -  \frac{\yqqb}{\yqbi}
    +  \frac{2\yqqb}{\yqi+\yqbi}
    \right)
\nonu
& & -2R'(-\yqqb,\yqi)\frac{\yqqb^2+(-\yqqb+\yqbi)^2}{\yqi\yqbi}
\nonu
& & -2R'(-\yqqb,\yqbi) \frac{\yqqb^2+(-\yqqb+\yqi)^2}{\yqi\yqbi}.
\end{eqnarray}

\noindent
The function $R'$ is given by
\begin{eqnarray}
   R'(x,y) & = & \ln\left|x\right|\ln\left|y\right|
                 \,-\,\ln\left|x\right|\ln\left|1-x\right|
                 \,-\,\ln\left|y\right|\ln\left|1-y\right|
\nonu
& & -\lim_{\eta\searrow 0}\mbox{Re}\left(\dilog(x+i\eta)+\dilog(y+i\eta)\right)
   + \zeta(2).
\end{eqnarray}

\section{Factorized Integration Kernels for Final State Singularities}
\label{kernelfss}

In this appendix the results for the singular kernels from the
final state singularities are summarised.
It is convenient to add additional indices to
the integration variables $r$, $z$ and $b$.
Let
\begin{equation}
z_j := \frac{p_0p_j}{p_0 q},
\end{equation}
where $p_0$ is the momentum of the incoming parton, and
\begin{equation}
r_{jk} := \frac{s_{jk}}{\SH y \xi}.
\end{equation}
If the integration  variable $b$ is fixed in the $(p_j,p_k)$ CM system,
let
\begin{equation}
b_j := \frac{1}{2}(1-\cos\theta_j), \quad
\theta_j:=\angle(\vec{p}_j, \vec{p}_0).
\end{equation}
Let $T_{q/g}(x_p,z_j)$ be the Born term with incoming quark/gluon,
expressed in the variables $x_p=\XB/\xi$ and $z_j$.
For the singular matrix elements one obtains (including the average over
colour degrees of freedom for incoming partons and the symmetry factor
for identical particles in the final state)
\begin{eqnarray}
\tr H_{F_1} & = & L_1 8\pi^2 \frac{\alpha_s}{2\pi} \mu^{2\epsilon}
                  16(1-\epsilon) C_F^2 Q_j^2 \frac{x_p}{Q^2}
\nonu &  &
                  \cdot
                  \frac{1}{r_{qg}}
                  \bigg[
                    (1-b_q)(1-\epsilon)-2+2\frac{1-z_g}{r_{qg}+(1-z_q)(1-b_q)}
                  \bigg]
                  T_q(x_p,z_g),
\\
\tr H_{F_2} & = & L_1 8\pi^2 \frac{\alpha_s}{2\pi} \mu^{2\epsilon}
                  (-16)(1-\epsilon) N_C C_F Q_j^2 \frac{x_p}{Q^2}
\nonu &  &
                  \cdot
                  \frac{1}{r_{qg}}
                  \bigg[
                    \frac{1-z_g}{r_{qg}+(1-z_g)(1-b_q)}
                   -\frac{1-x_p-r_{qg}}{r_{qg}+(1-x_p-r_{qg})(1-b_q)}
                  \bigg]
                  T_q(x_p,z_g),
\\
\tr H_{F_3} & = & L_1 8\pi^2 \frac{\alpha_s}{2\pi} \mu^{2\epsilon}
                  8(1-\epsilon) N_C C_F Q_j^2 \frac{x_p}{Q^2}
\nonu &  &
                  \cdot
                  \frac{1}{r_{gg}}
                  \bigg[
                    \frac{1-x_p-r_{gg}}{r_{gg}+(1-x_p-r_{gg})b_g}
                   +\frac{1-z_q}{r_{gg}+(1-z_q)b_g}
\nonu &  &
                   +\frac{1-x_p-r_{gg}}{r_{gg}+(1-x_p-r_{gg})(1-b_g)}
                   +\frac{1-z_q}{r_{gg}+(1-z_q)(1-b_g)}-2
                  \bigg]
                  T_q(x_p,z_q),
\\
\tr H_{F_4} & = & L_1 8\pi^2 \frac{\alpha_s}{2\pi} \mu^{2\epsilon}
                  (-16)(1-\epsilon) N_C C_F Q_j^2 \frac{x_p}{Q^2}
                  \frac{1}{r_{gg}}
                  \bigg[
                    1-b_g+b_g^2
                  \bigg]
                  T_q(x_p,z_q)
\nonu &  &
                  + \mbox{terms}\sim \left(1-2(1-\epsilon)\cos^2\varphi\right),
\\
\tr H_{F_5} & = & L_1 8\pi^2 \frac{\alpha_s}{2\pi} \mu^{2\epsilon}
                  8(1-\epsilon) C_F Q_j^2 \frac{x_p}{Q^2}
\nonu &  &
                  \cdot
                  \frac{1}{r_{q\iqbar}}
                  \bigg[
                    1-2b_q(1-b_q)(1+\epsilon)
                  \bigg]
                  T_q(x_p,z_q)
\nonu &  &
                  + \mbox{terms}\sim \left(1-2(1-\epsilon)\cos^2\varphi\right),
\\
\tr H_{F_6} & = & L_1 8\pi^2 \frac{\alpha_s}{2\pi} \mu^{2\epsilon}
                  8(1-\epsilon) C_F Q_j^2 \frac{x_p}{Q^2}
\nonu &  &
                  \cdot
                  \frac{1}{r_{qg}}
                  \bigg[
                    (1-b_q)(1-\epsilon)-2
                   +2\frac{1-x_p-r_{qg}}{r_{qg}+(1-x_p-r_{qg})(1-b_q)}
                  \bigg]
                  T_g(x_p,z_{\iqbar})
\nonu &  &
                  + (q \leftrightarrow \qbar),
\\
\tr H_{F_7} & = & L_1 8\pi^2 \frac{\alpha_s}{2\pi} \mu^{2\epsilon}
                  (-8)(1-\epsilon) N_C Q_j^2 \frac{x_p}{Q^2}
\nonu &  &
                  \cdot
                  \frac{1}{r_{qg}}
                  \bigg[
                    \frac{1-x_p-r_{qg}}{r_{qg}+(1-x_p-r_{qg})(1-b_q)}
                   -\frac{1-z_{\iqbar}}{r_{qg}+(1-z_{\iqbar})(1-b_q)}
                  \bigg]
                  T_g(x_p,z_{\iqbar})
\nonu &  &
                  + (q \leftrightarrow \qbar).
\end{eqnarray}
In this factorization terms that vanish
(after the integration)
for $c\rightarrow 0$ are neglected.
Therefore the cut should not be too large in
the numerical evaluation. For small values of the cut the cross section is
dominated by terms $\sim\left(-\ln^2c\right)$.

\section{Phase Space Integrals for Final State Singularities}
\label{appri}
\label{appc}

In this appendix the results for the real corrections
of the terms involving final state singularities are collected.
A measure $d\mu_F$ is defined by
\begin{equation}
\int d\mu_F:=\int_0^\alpha dr r^{-\epsilon}
            \left(1-\frac{r}{h}
            \right)^{-\epsilon}
            \int_0^1 \frac{db}{N_b} b^{-\epsilon} (1-b)^{-\epsilon}
            \int_o^\pi \frac{d\varphi}{N_\varphi} \sin^{-2\epsilon}\varphi.
\end{equation}
Integrals of terms with singularities for $r \rightarrow 0$,
$b \rightarrow 0$ are needed in the integrations over the singular
region of phase space. $N_b$ and $N_\varphi$ are normalisation constants
from the phase space, and $h$ is an arbitrary parameter which does not
show up in the results up to order \porder{\epsilon^0}. The upper limit
$\alpha$ of the $r$-integration has the meaning of a jet cut.
The integrals that were needed were solved by use of
\cite{37,38,39,40} and are of the following type:

\begin{eqnarray}
f_1(y) & := & \int d\mu_F \frac{1}{r} \frac{y}{r+y b}
\nonu
       & = & \frac{1}{2\epsilon^2}
           +\frac{1}{\epsilon}\left(-1-\frac{1}{2}\ln(y)\right)
           +\ln y - \frac{1}{2}\ln^2\frac{\alpha}{y}
           +\frac{1}{4}\ln^2y
\nonu
       &   & \quad {}-\spence\left(-\frac{\alpha}{y}\right)-\zeta(2)
           +\porder{\epsilon},
\\
f_2(y) & := & \int d\mu_F \frac{1}{r} \frac{y-r}{r+(y-r) b}
\nonu
       & = & \frac{1}{2\epsilon^2}
           +\frac{1}{\epsilon}\left(-1-\frac{1}{2}\ln(y)\right)
           +\ln y - \frac{1}{2}\ln^2\frac{\alpha}{y}
           +\frac{1}{4}\ln^2y-\zeta(2)
           +\porder{\epsilon},
\\
f_3 & := & \int d\mu_F \frac{1}{r} (1-b)(1-\epsilon)
\nonu
       & = & -\frac{1}{2\epsilon}+\frac{1}{2}
           +\frac{1}{2}\ln\alpha
           +\porder{\epsilon},
\\
f_4 & := & \int d\mu_F \frac{1}{r}
\nonu
       & = & -\frac{1}{\epsilon}
           +\ln\alpha
           +\porder{\epsilon},
\\
f_5 & := & \int d\mu_F \frac{1}{r} (1-b+b^2)
\nonu
       & = & -\frac{5}{6}\frac{1}{\epsilon}
           +\frac{5}{6}\ln\alpha-\frac{1}{18}
           +\porder{\epsilon},
\\
f_6 & := & \int d\mu_F \frac{1}{r} b(1-b)(1+\epsilon)
\nonu
       & = & -\frac{1}{6}\frac{1}{\epsilon}
           -\frac{1}{9}+\frac{1}{6}\ln\alpha
           +\porder{\epsilon}.
\end{eqnarray}
The function $f_1(y)$ can be checked against the result in \cite{27}.

\section{Real Corrections, Final State Singularities}
\label{appd}

By means of the basic integrals from appendix \ref{appri}
one obtains the explicit expressions
for the final state real corrections:

\begin{eqnarray}
F_1 & = & C_F^2 Q_j^2 T_q(x_p,z_g)
\nonu
       &  & \quad\cdot\bigg\{
             \frac{1}{\epsilon^2}
             +\frac{1}{\epsilon}\left(\frac{3}{2}-\ln\frac{1-z_g}{x_p}\right)
             +\frac{7}{2}-\frac{3}{2}\ln\frac{\alpha}{x_p}
             -\ln x_p \ln (1-z_g)
\nonu
       &  & \quad {}+\frac{1}{2}\ln^2 x_p -\ln^2\frac{\alpha}{1-z_g}
           +\frac{1}{2}\ln^2(1-z_g)-2\spence\left(\frac{\alpha}{1-z_g}\right)
           -2\zeta(2)
           \bigg\}+\porder{\epsilon},
\\
F_2 & = & -\frac{1}{2} N_C C_F Q_j^2 T_q(x_p,z_g)
\nonu
       &  & \quad\cdot\bigg\{
            \frac{1}{\epsilon}\ln\frac{1-x_p}{1-z_g}
           +\ln x_p \ln\frac{1-x_p}{1-z_g} + \ln^2\frac{\alpha}{1-x_p}
           -\ln^2\frac{\alpha}{1-z_g}
\nonu
       &  & +\frac{1}{2}\left(
            \ln^2(1-z_g)-\ln^2(1-x_p)
           \right)
            -2\spence\left(-\frac{\alpha}{1-z_g}\right)
           \bigg\}+\porder{\epsilon},
\\
F_3 & = & -\frac{1}{2} N_C C_F Q_j^2 T_q(x_p,z_q)
\nonu
       &  & \quad\cdot\bigg\{
            -\frac{2}{\epsilon^2}
            +\frac{1}{\epsilon}\left(-2-\ln\frac{x_p^2}{(1-z_q)(1-x_p)}\right)
            -4+2\ln\frac{\alpha}{x_p}
\nonu
       &  & +\ln x_p\ln\left((1-z_g)(1-x_p)\right)
            -\ln^2x_p+\ln^2\frac{\alpha}{1-x_p}
            +\ln^2\frac{\alpha}{1-z_q}
\nonu
       &  & -\frac{1}{2}\left(\ln^2(1-x_p)+\ln^2(1-z_q)\right)
            +4\zeta(2)+2\spence\left(-\frac{\alpha}{1-z_q}\right)
            \bigg\}+\porder{\epsilon},
\\
F_4 & = & -\frac{1}{2} N_C C_F Q_j^2 T_q(x_p,z_q)
\nonu
       &  & \quad\cdot\bigg\{
            -\frac{5}{3}\frac{1}{\epsilon}+\frac{5}{3}\ln\frac{\alpha}{x_p}
            -\frac{31}{9}
            \bigg\}+\porder{\epsilon},
\\
F_5 & = & C_F Q_j^2 T_q(x_p,z_q)
\nonu
       &  & \quad\cdot\bigg\{
            -\frac{1}{3}\frac{1}{\epsilon}-\frac{5}{9}
            +\frac{1}{3}\ln\frac{\alpha}{x_p}
            \bigg\}+\porder{\epsilon},
\\
F_6 & = & \frac{1}{2} C_F Q_j^2 T_g(x_p,z_{\iqbar})
\nonu
       &  & \quad\cdot\bigg\{
            \frac{1}{\epsilon^2}
            +\frac{1}{\epsilon}\left(\frac{3}{2}-\ln\frac{1-x_p}{x_p}\right)
            +\frac{7}{2}-\frac{3}{2}\ln\frac{\alpha}{x_p}-\ln x_p\ln(1-x_p)
\nonu
       &  & +\frac{1}{2}\ln^2x_p+\frac{1}{2}\ln^2(1-x_p)
            -\ln^2\frac{\alpha}{1-x_p}-2\zeta(2)
            \bigg\}+\porder{\epsilon}
\nonu
       &  & +(q \longleftrightarrow {\qbar}),
\\
F_7 & = & -\frac{1}{4} N_C Q_j^2 T_g(x_p,z_{\iqbar})
\nonu
       &  & \quad\cdot\bigg\{
            \frac{1}{\epsilon}\ln\frac{1-z_{\iqbar}}{1-x_p}
            +\ln x_p\ln\frac{1-z_{\iqbar}}{1-x_p}
            +\ln^2\frac{\alpha}{1-z_{\iqbar}}
            -\ln^2\frac{\alpha}{1-x_p}
\nonu
       &  & +\frac{1}{2}\left(\ln^2(1-x_p)
            -\ln^2(1-z_{\iqbar})\right)
            +2\spence\left(-\frac{\alpha}
            {1-z_{\iqbar}}\right)
            \bigg\}+\porder{\epsilon}
\nonu
       &  & +(q \longleftrightarrow {\qbar}).
\end{eqnarray}

\section{Factorized Integration Kernels for Initial State Singularities}
\label{kerneliss}

This section contains the singular kernels from the factorization
of the matrix elements with initial state singularities.
Let $T_{q/g}(t_{jk},z_l)$ be the Born term with incoming quark/gluon,
expressed in the variables $t_{jk}$ and $z_l$.
The singular matrix elements then read
\begin{eqnarray}
\tr H_{I_1} & = & L_1 8\pi^2 \frac{\alpha_s}{2\pi} \mu^{2\epsilon}
                  16(1-\epsilon) C_F^2 \frac{1}{W^2}Q_j^2
\nonu &  &
                  \cdot
                  \frac{1}{z'}
                  \bigg[
                  \left(
                    \frac{1}{-(1-\nu-t_{qg})}
                   -\frac{1}{-(1-\nu-t_{qg})+(1-t_{qg})\sigma}
                  \right)(1-\epsilon)
\nonu & &
                  +2\frac{z_q}{-(1-\nu-t_{qg})z'+(1-t_{qg})z_q\sigma}
                  \bigg]
                  T_q(t_{qg},z_q),
\\
\tr H_{I_2} & = & L_1 8\pi^2 \frac{\alpha_s}{2\pi} \mu^{2\epsilon}
                  (-16)(1-\epsilon) N_C C_F \frac{1}{W^2}Q_j^2
\nonu &  &
                  \cdot
                  \frac{1}{z'}
                  \bigg[
                  \frac{z_q}{-(1-\nu-t_{qg})z'+(1-t_{qg})z_q\sigma}
\nonu & &
                  -\frac{1-z_q}{-(1-\nu-t_{qg})z'+(1-t_{qg})(1-z_q)\sigma}
                  \bigg]
                  T_q(t_{qg},z_q),
\\
\tr H_{I_3} & = & L_1 8\pi^2 \frac{\alpha_s}{2\pi} \mu^{2\epsilon}Q_j^2
\nonu & &
                  \cdot
                  \bigg\{
                  8(1-\epsilon) C_F \frac{1}{W^2}T_g(t_{q\iqbar},z_q)
                  (1-\epsilon)
\nonu & &
                  \cdot
                  \frac{1}{z'}
                  \bigg[
                  \frac{1}{-(1-\nu-t_{q\iqbar})+(1-t_{q\iqbar})\sigma}
                  +\frac{2(1-t_{q\iqbar})\sigma}{(1-\nu-t_{q\iqbar})^2}
                   (1+\epsilon)
                  \bigg]
\nonu & &
                  -16 C_F \frac{1}{W^2}\frac{1}{z'}
                  \frac{2(1-\nu)t_{q\iqbar}\sigma(1-t_{q\iqbar})}
                       {(1-\nu-t_{q\iqbar})^4z_q(1-z_q)}
                  \left(1-2(1-\epsilon)\cos^2\varphi\right)
                  \bigg\},
\\
\tr H_{I_4} & = & L_1 8\pi^2 \frac{\alpha_s}{2\pi} \mu^{2\epsilon}
                  8(1-\epsilon) C_F \frac{1}{W^2}Q_j^2
                  \frac{1}{1-\epsilon}
\nonu &  &
                  \cdot
                  \frac{1}{z'}
                  \bigg[
                  \frac{1}{-(1-\nu-t_{qg})}(1-\epsilon)
                  -2\frac{\sigma(1-t_{qg})}
                         {\left(-(1-\nu-t_{qg})
                         +(1-t_{qg})\sigma\right)^2}
                  \bigg]
                  T_q(t_{qg},z_q)
\nonu &  &
                  + (q \leftrightarrow \qbar),
\\
\tr H_{I_5} & = & 0,
\\
\tr H_{I_6} & = & L_1 8\pi^2 \frac{\alpha_s}{2\pi} \mu^{2\epsilon}
                  8(1-\epsilon) N_C \frac{1}{W^2}Q_j^2
\nonu &  &
                  \cdot
                  \frac{1}{z'}
                  \bigg[
                  \frac{1-z_q}{-(1-\nu-t_{q\iqbar})z'
                              +(1-t_{q\iqbar})(1-z_q)\sigma}
                  +\frac{z_q}{-(1-\nu-t_{q\iqbar})z'
                              +(1-t_{q\iqbar})z_q\sigma}
\nonu & &
                  -2\frac{-(1-\nu-t_{q\iqbar})}
                         {\left(-(1-\nu-t_{q\iqbar})
                         +(1-t_{q\iqbar})\sigma\right)^2}
                  \bigg]
                  T_g(t_{q\iqbar},z_q),
\\
\tr H_{I_7} & = & L_1 8\pi^2 \frac{\alpha_s}{2\pi} \mu^{2\epsilon}Q_j^2
\nonu & &
                  \cdot
                  \bigg\{
                  16(1-\epsilon) N_C \frac{1}{W^2}T_g(t_{q\iqbar},z_q)
                  \frac{1}{z'}
                  \bigg[
                  \frac{1}{-(1-\nu-t_{q\iqbar})+(1-t_{q\iqbar})\sigma}
                  +\frac{(1-t_{q\iqbar})\sigma}{(1-\nu-t_{q\iqbar})^2}
                  \bigg]
\nonu & &
                  -16 N_C \frac{1}{W^2}\frac{1}{z'}
                  \frac{2t_{q\iqbar}(1-\nu)(1-\zeta-t_{q\iqbar})}
                       {(1-\nu-t_{q\iqbar})^4z_q(1-z_q)}
                  \left(1-2(1-\epsilon)\cos^2\varphi\right)
                  \bigg\}.
\end{eqnarray}

In these formulae the terms proportional
to $1-2(1-\epsilon)\cos^2\varphi$ are stated explicitly,
since for invariant mass cuts
the integration over the azimuthal angle is (in general) not over the
range $[0,\pi]$, and as a consequence
\begin{equation}
 \int d\varphi \sin^{-2\epsilon}\varphi (1-2(1-\epsilon)\cos^2\varphi)
\end{equation}
does not vanish.\footnote{These terms therefore
do not factorize in the familar form of a singular kernel
multiplied by the Born term. However, this does {\em not} affect the
factorization of the singularities because the integration region for
$\varphi$ is $[0,\pi]$ in the limit $z'\rightarrow 0$. Therefore the resulting
contribution is not divergent although the integrand is singular in this
limit.}
Additional factors of
$(1-\epsilon)$ and $1/(1-\epsilon)$ in $\tr H_{I_3}$ and
$\tr H_{I_4}$, respectively, are due to the fact that in $d$ dimensions
a gluon has $(d-2)$ (physical) helicity states, whereas a quark
has only $2$, and the incoming particle in the (3+1) jet Born term
is not the same as that in the factorized (2+1) jet Born term.

\section{Phase Space Integrals for Initial State Singularities}
\label{appe}

Here the results for the integrals needed
for the real corrections with initial state singularities are stated.
A measure $d\mu_I$ can be defined by
\begin{equation}
\int d\mu_I:=\int_0^\beta dz' z'^{-\epsilon}
             \int_0^1 d\sigma \sigma^{-\epsilon}
             \int_0^\pi \frac{d\varphi}{N_\varphi} \sin^{-2\epsilon} \varphi
             \frac{\Gamma(1-2\epsilon)}{\Gamma^2(1-\epsilon)}
             a \left(\frac{1-\XB}{\XB}\right)^{-\epsilon}
             (1-t)^{1-\epsilon}.
\end{equation}
For an arbitrary $C^\infty$-function $g:{[}0,1{]}\rightarrow\cnumber$
let
\begin{equation}
   F(\sigma){[}g{]}:=\int_0^1d\sigma F(\sigma) g(\sigma).
\end{equation}
Then the following integrals are given by
\begin{eqnarray}
i_1(y) & := & \int d\mu_I g(\sigma)\frac{1}{z'} \frac{y}{z'+y\sigma}
\nonu
       & = & a(1-t)\bigg[
            \bigg\{
            \frac{1}{2\epsilon^2}
           +\frac{1}{\epsilon}\left(-\frac{1}{2}\ln\frac{(1-a)y}{\XB}\right)
           +\frac{1}{4}\ln^2\frac{(1-a)y}{\XB}
\nonu & &
           -\frac{1}{2}\ln^2\frac{\beta}{y}
           -\spence\left(-\frac{\beta}{y}\right)
            \bigg\}\delta_0{[}g{]}
\nonu
       & & +\bigg\{
            -\frac{1}{\epsilon}
           +\ln\frac{(1-a)y}{\XB}
            \bigg\}\left(\frac{1}{\sigma}\right)_+{[}g{]}
           +2\left(\frac{\ln\sigma}{\sigma}\right)_+{[}g{]}
\nonu & &
           -\left(\frac{\ln\left(1+\frac{\sigma y}{\beta}\right)}{\sigma}
            \right)_+{[}g{]}
           \bigg]+\porder{\epsilon},
\\
i_2    & := & \int d\mu_I g(\sigma)\frac{1}{z'}
\nonu
       & = & a(1-t)\bigg[
            -\frac{1}{\epsilon}\munit{[}g{]}
            +\left( \ln(\frac{(1-a)\beta\sigma}{\XB}\right){[}g{]}
           \bigg]+\porder{\epsilon},
\\
i_3(y) & := & \int d\mu_I g(\sigma)\frac{1}{z'}\frac{y}{1+y\sigma}
\nonu
       & = & a(1-t)\bigg[
            -\frac{1}{\epsilon}\left(\frac{y}{1+y\sigma}\right){[}g{]}
            +\left(\frac{y}{1+y\sigma}
                   \ln(\frac{(1-a)\beta\sigma}{\XB}\right){[}g{]}
           \bigg]+\porder{\epsilon},
\\
i_4    & := & \int d\mu_I g(\sigma)\frac{1}{z'}\sigma
\nonu
       & = & a(1-t)\bigg[
            -\frac{1}{\epsilon}\sigma{[}g{]}
            +\left(\sigma
                   \ln(\frac{(1-a)\beta\sigma}{\XB}\right){[}g{]}
           \bigg]+\porder{\epsilon},
\\
i_5(y) & := & \int d\mu_I g(\sigma)\frac{1}{z'}
              \frac{y^2\sigma}{(1+y\sigma)^2}
\nonu
       & = & a(1-t)\bigg[
            -\frac{1}{\epsilon}\left(
                      \frac{y^2\sigma}{(1+y\sigma)^2}\right){[}g{]}
            +\left(\frac{y^2\sigma}{(1+y\sigma)^2}
                   \ln\frac{(1-a)\beta\sigma}{\XB}\right){[}g{]}
           \bigg]+\porder{\epsilon},
\\
i_6(y) & := & \int d\mu_I g(\sigma)\frac{1}{z'}
              \frac{y^2}{(1+y\sigma)^2}
\nonu
       & = & a(1-t)\bigg[
            -\frac{1}{\epsilon}\left(
                      \frac{y^2}{(1+y\sigma)^2}\right){[}g{]}
            +\left(\frac{y^2}{(1+y\sigma)^2}
                   \ln\frac{(1-a)\beta\sigma}{\XB}\right){[}g{]}
           \bigg]+\porder{\epsilon}.
\end{eqnarray}

\noindent
Here the distributions
\begin{equation}
   \delta_0{[}g{]}:=g(0), \quad
   \munit{[}g{]}:=\int_0^1d\sigma g(\sigma)
\end{equation}
have been used.

\noindent
The ``+"-prescription for the $\sigma$-integration is defined by
\begin{equation}
   \int_0^1d\sigma D_+(\sigma)g(\sigma)
  :=\int_0^1d\sigma D(\sigma) \left(g(\sigma)-g(0)\right).
\end{equation}

\section{Real Corrections, Initial State Singularities}
\label{appf}

The explicit expressions of the sum of
the (2+1) and (3+1) jet contributions from the
initial state singularities read
\begin{eqnarray}
I_1 & = & Q_j^2 T_q(t_{qg},z_q)
\nonu
       &  & \quad\cdot\bigg\{
             \left(\frac{Q^2}{M_f^2}\right)^{\epsilon}
             \left(-\frac{1}{\epsilon}+\log\frac{Q^2}{M_f^2}
             \right) P_{q\leftarrow q}(u) C_F
\nonu
       &  &  +C_F^2\bigg{[}
                   \frac{1}{\epsilon^2}\delta(1-u)
                  +\frac{1}{\epsilon}\left(
                     -\ln\frac{az_q}{\XB}\delta(1-u)+\frac{3}{2}\delta(1-u)
                                     \right)
\nonu
       &  &       +S_1\left(\frac{a\beta}{\XB},
                            \frac{az_q}{\XB},
                            \frac{z_q}{\beta}
                      \right)
                   \bigg{]}
           \bigg\}+\porder{\epsilon},
\\
I_2 & = & Q_j^2 T_q(t_{qg},z_q)\cdot\left(-\frac{1}{2}N_C C_F\right)
\nonu
       &  & \quad\cdot\bigg\{
             \frac{1}{\epsilon}\ln\frac{1-z_q}{z_q}\delta(1-u)
             +S_2\left(
                 \frac{az_q}{\XB},\frac{a(1-z_q)}{\XB},\frac{z_q}{\beta},
                 \frac{1-z_q}{\beta}
                 \right)
           \bigg\}+\porder{\epsilon},
\\
I_3 & = & Q_j^2 T_g(t_{q{\iqbar}},z_q)
\nonu
       &  & \quad\cdot\bigg\{
             \left(\frac{Q^2}{M_f^2}\right)^{\epsilon}
             \left(
             -\frac{1}{\epsilon}+\log\frac{Q^2}{M_f^2}+1
             \right)P_{g\leftarrow q}(u)\cdot\frac{1}{2}
\nonu
       &  & +\frac{1}{2}C_F S_3\left(\frac{a\beta}{\XB}\right)
           \bigg\}+\porder{\epsilon},
\\
I_4 & = & Q_j^2 T_q(t_{qg},z_q)
\nonu
       &  & \quad\cdot\bigg\{
             \left(\frac{Q^2}{M_f^2}\right)^{\epsilon}
             \left(-\frac{1}{\epsilon}+\log\frac{Q^2}{M_f^2}-1
             \right)P_{q\leftarrow g}(u) C_F
\nonu
       &  & +C_F S_4\left(\frac{a\beta}{\XB}\right)
           \bigg\}
           +(q \longleftrightarrow {\qbar})
           +\porder{\epsilon},
\\
I_5 & = & \porder{\epsilon},
\\
I_6+I_7 & = & Q_j^2 T_g(t_{q{\iqbar}},z_q)
\nonu
       &  & \quad\cdot\bigg\{
             \left(\frac{Q^2}{M_f^2}\right)^{\epsilon}
             \left(-\frac{1}{\epsilon}+\log\frac{Q^2}{M_f^2}
             \right)P_{g\leftarrow g}(u) \cdot\frac{1}{2}
\nonu
       &  &  +\left(-\frac{1}{4}N_C\right)\bigg{[}
                  -\frac{2}{\epsilon^2}\delta(1-u)
\nonu  &  &
                  +\frac{1}{\epsilon}\left(
                     \ln\frac{a^2 z_q(1-z_q)}{\XB^2}\delta(1-u)
                     -\frac{2}{N_C}\left(\frac{11}{6}N_C-\frac{1}{3}N_F
                    \right)\delta(1-u)\right)
\nonu
       &  &       +S_6\left(\frac{az_q}{\XB},
                            \frac{a(1-z_q)}{\XB},
                            \frac{z_q}{\beta},
                            \frac{1-z_q}{\beta},
                            \frac{a\beta}{\XB}
                      \right)
                   \bigg{]}
           \bigg\}+\porder{\epsilon}.
\end{eqnarray}
The functions $P_{B\leftarrow A}$ are the Altarelli-Parisi kernels
and the functions $S_i$ are defined by
\begin{eqnarray}
S_1(A,B,C) & = & R_2+R_5(A)-R_6(A)
                +\left(\frac{1}{2}\ln^2 B+6\zeta(2)\right)R_1
\nonu & &
                +2R_3(B)-2R_4(C),
\\
S_2(A,B,C,D) & = & R_1\left(\frac{1}{2}\ln^2A-\frac{1}{2}\ln^2B\right)
                  +2\left(R_3(A)-R_3(B)\right)
\nonu & &
                  -2\left(R_4(C)-R_4(D)\right),
\\
S_3(A) & = & R_6(A)+2R_7(A)-2R_{10},
\\
S_4(A) & = & \frac{1}{2} R_0 +\frac{1}{2} R_5(A)-R_8(A),
\\
S_6(A,B,C,D,E) & = & -R_1\left(\frac{1}{2}\ln^2A
                              +\frac{1}{2}\ln^2B+12\zeta(2)\right)
                     -2\left(R_3(A)+R_3(B)\right)
\nonu & &
                     +2\left(R_4(C)+R_4(D)\right)
                     +4R_9(E)-4R_6(E)-4R_7(E),
\end{eqnarray}
and the functions $R_i$ are distributions in the variable $u$ given by
\begin{eqnarray}
R_0 & = & 1,
\\
R_1 & = & \delta(1-u),
\\
R_2 & = & 1-u,
\\
R_3(\lambda) & = & \left(\frac{\ln\left[\lambda\left(
                                 \frac{1-u}{u}\right)^2
                                 \right]}{1-u} \right)_+
                  -\ln\left(\lambda\left(\frac{1-u}{u}\right)^2 \right),
\\
R_4(\lambda) & = & \frac{u}{1-u}\ln\left(1+\lambda\frac{1-u}{u}\right),
\\
R_5(\lambda) & = & \ln\left(\lambda\frac{1-u}{u}\right),
\\
R_6(\lambda) & = & u \ln\left(\lambda\frac{1-u}{u}\right),
\\
R_7(\lambda) & = & \frac{1-u}{u} \ln\left(\lambda\frac{1-u}{u}\right),
\\
R_8(\lambda) & = & u(1-u) \ln\left(\lambda\frac{1-u}{u}\right),
\\
R_9(\lambda) & = & u^2 \ln\left(\lambda\frac{1-u}{u}\right),
\\
R_{10}(\lambda) & = & \frac{1-u}{u}.
\end{eqnarray}

\end{appendix}

\newpage
%
%
\newcommand{\bibitema}[1]{\bibitem[#1]{#1}}

\newpage
\noindent
{\Large \bf Figure
Captions:}

\begin{description}
\item[Fig.~1] Diagram for initial state radiation.
\item[Fig.~2] Diagram for 1-parton production.
\item[Fig.~3] Diagram for the virtual correction to 1-parton production
\item[Fig.~4] Generic diagrams for 2-parton production.
\item[Fig.~5] Born terms for 2-parton production (hadron tensor).
\item[Fig.~6] Virtual corrections to 2-parton production (hadron tensor).
\item[Fig.~7] Diagrams contributing to the wave function renormalisation.
\item[Fig.~8] Born terms of $\porder{\alpha_s^2}$
(the roman numbering labels the 7 different colour classes).
\item[Fig.~9] Generic diagrams for 3-parton production.
\item[Fig.~10] CM frame of $p_1$ and $p_2$.
\item[Fig.~11] (2+1) and (3+1) jet regions in phase space.
\item[Fig.~12] Cut dependence of jet cross sections. ``{\em tr.}'' stands for
transverse, ``{\em long.}'' for longitudinal polarisation of the exchanged
virtual photon. Contributions marked as ``{\em NLO}'' are
given by the sum of the Born term
and the next-to-leading order contribution.
(a)--(c) HERA kinematics with
(a) 3.16~\GeV~$<$~$Q$~$<$~10~\GeV,
(b) 10~\GeV~$<$~$Q$~$<$~31.6~\GeV,
(c) 31.6~\GeV~$<$~$Q$~$<~$100~\GeV.
(d) is the graph for the E665 kinematics.
\item[Fig.~13] The ratio
$\sigma_{\mbox{\small 2+1}}/\sigma_{\mbox{\small tot}}$.
(a)--(c) HERA kinematics with
(a) 3.16~\GeV~$<$~$Q$~$<$~10~\GeV,
(b) 10~\GeV~$<$~$Q$~$<$~31.6~\GeV,
(c) 31.6~\GeV~$<$~$Q$~$<$~100~\GeV,
(d) E665 kinematics.
\item[Fig.~14] Dependence of the cross section on the jet definition scheme.
The jet definition scheme is
(a) $s_{ij}\grkl cW^2$,
(b) $s_{ij}\grkl c\left(W^\alpha Q^\beta
\sqrt{\SH y}^{1-\alpha-\beta}\right)^2$
with
$\alpha=0.8$, $\beta=0.2$.
\item[Fig.~15] Dependence on the renormalisation and factorization scales.
(a) renormalisation scale $\mu=\rho Q$, factorization scale $M_f=Q$;
(b) $\mu=Q$, $M_f=\rho Q$;
(c) $\mu=\rho Q$, $M_f=\rho Q$.
(d)--(f) are the same, but for much larger Q.
(g)--(i) are the graphs for E665 energies.
To avoid scales that are too low for perturbation theory to be valid,
the scales are clipped at a lower bound of
2~\GeV{} (HERA) and 1~\GeV{} (E665).
\item[Fig.~16] Dependence of the jet rate
$\sigma_{\mbox{\small 2+1}}/\sigma_{\mbox{\small tot}}$
on the renormalisation and factorization scales.
The indices (a)--(i) refer to the indices of Fig. \ref{fig15}
(a)--(i), respectively.
\item[Fig.~17] Dependence on the parton densities.
The plotted cross section is $\XB\,d\sigma/d\XB$, the jet definition scheme is
(a) $s_{ij}\grkl cW^2$,
(b) $s_{ij}\grkl c\left(W^\alpha Q^\beta
\sqrt{\SH y}^{1-\alpha-\beta}\right)^2$
with
$\alpha=0.7$, $\beta=0.3$.
\end{description}

\newpage
\setcounter{totalnumber}{20}

\begin{figure}[h]
  \vspace{5cm}
  \caption{\label{irad1}}
\end{figure}

\begin{figure}[h]
  \vspace{5cm}
  \caption{\label{j1}}
\end{figure}

\begin{figure}[h]
  \vspace{5cm}
  \caption{\label{j2}}
\end{figure}

\begin{figure}[h]
  \vspace{5cm}
  \caption{\label{diag1}}
\end{figure}

\begin{figure}[h]
  \vspace{4cm}
  \caption{\label{diag2}}
\end{figure}

\begin{figure}[h]
  \vspace{6cm}
  \caption{\label{diag3}}
\end{figure}

\begin{figure}[h]
  \vspace{7cm}
  \caption{\label{diag4}}
\end{figure}

\begin{figure}[h]
  \vspace{7cm}
  \caption{\label{diag41}}
\end{figure}

\begin{figure}[h]
  \vspace{5.5cm}
  \caption{\label{diag42}}
\end{figure}

\begin{figure}[h]
  \vspace{5cm}
  \caption{\label{diag44}}
\end{figure}

\begin{figure}[h]
  \vspace{5cm}
  \caption{\label{diag53}}
\end{figure}

\clearpage
\newpage

\begin{figure}[h]
  \vspace{0cm}
  \caption{\label{fig12}}
\end{figure}

\begin{figure}[h]
  \vspace{0cm}
  \caption{\label{fig13}}
\end{figure}

\begin{figure}[h]
  \vspace{0cm}
  \caption{\label{fig14}}
\end{figure}

\begin{figure}[h]
  \vspace{0cm}
  \caption{\label{fig15}}
\end{figure}

\begin{figure}[h]
  \vspace{0cm}
  \caption{\label{fig16new}}
\end{figure}

\begin{figure}[h]
  \vspace{0cm}
  \caption{\label{fig16}}
\end{figure}


\end{document}